\documentclass[11pt,english,superscriptaddress,nofootinbib,preprintnumbers, aps]{revtex4}

\usepackage{amsmath,amsthm,latexsym,amssymb,amsfonts,epsfig}

\usepackage{float}
\usepackage{upgreek}

\theoremstyle{remark}
\newtheorem{remark}{Remark}[section]
\theoremstyle{remark}

\theoremstyle{obser}
\newtheorem{obser}{Observation}[section]
\theoremstyle{obser}

\usepackage{xcolor}
\usepackage{amsmath}
\usepackage{amsthm}
\usepackage{graphicx}
\usepackage{amssymb}
\usepackage{dsfont}
\usepackage{color}
\usepackage{soul}
\usepackage[hyperfootnotes=true,unicode=true, 
 bookmarks=true,bookmarksnumbered=false,bookmarksopen=false,
 breaklinks=false,pdfborder={0 0 1},backref=false,colorlinks=true]
 {hyperref}
 \usepackage{float}
 \usepackage[markup=nocolor]{changes}

\usepackage[makeroom]{cancel}

\usepackage{etoolbox}

\usepackage{hyperref}
\hypersetup{
    colorlinks,%
    citecolor=blue,%
    filecolor=blue,%
    linkcolor=blue,%
    urlcolor=blue
}

\newtheorem{Theorem}{Theorem}
\newtheorem{Definition}{Definition}

\newcommand{\be}{\begin{equation}}
\newcommand{\ee}{\end{equation}}
\newcommand{\ba}{\begin{eqnarray}}
\newcommand{\ea}{\end{eqnarray}}
\def\bal#1\eal{\begin{align}#1\end{align}}

\makeatletter
\@addtoreset{equation}{section}
\makeatletter
\@ifundefined{textcolor}{}
{%
 \definecolor{BLACK}{gray}{0}
 \definecolor{WHITE}{gray}{1}
 \definecolor{RED}{rgb}{1,0,0}
 \definecolor{GREEN}{RGB}{0,204,0}
 \definecolor{BLUE}{rgb}{0,0,1}
 \definecolor{CYAN}{cmyk}{1,0,0,0}
 \definecolor{MAGENTA}{cmyk}{0,1,0,0}
 \definecolor{YELLOW}{cmyk}{0,0,1,0}
 }

\usepackage{hyperref}
\hypersetup{
    colorlinks,%
    citecolor=blue,%
    filecolor=blue,%
    linkcolor=blue,%
    urlcolor=blue
}

\makeatother

\newcommand{\re}{\mathbb{R}}
\newcommand{\Hil}{\mathcal{H}}
\newcommand{\id}{\mathds{1}}

\begin{document}

\title{Algorithmic approach to Cosmological Coherent State Expectation Values in LQG}

\author{Klaus Liegener}
\email{klaus.liegener@tum.de}
\affiliation{Institute for Quantum Gravity, Friedrich-Alexander University Erlangen-N\"urnberg, Staudtstra\ss e 7, 91058 Erlangen, Germany}
\affiliation {II. Institute for Theoretical Physics, University of Hamburg, Luruper Chaussee 149, 22761 Hamburg, Germany}

\author{\L ukasz Rudnicki} 
\email{lukasz.rudnicki@ug.edu.pl}
\affiliation{International Centre for Theory of Quantum Technologies (ICTQT), University of Gda{\'n}sk, 80-308 Gda{\'n}sk, Poland}
\affiliation{Center for Theoretical Physics, Polish Academy of Sciences, Al. Lotnik{\'o}w 32/46, 02-668 Warsaw, Poland}

\date{\today{}}

\begin{abstract}
\fontfamily{lmss}\selectfont{
In the lattice approach to Loop Quantum Gravity on a fixed graph computations tend to be involved and are rarely analytically manageable. But, when interested in the expectation values of coherent states on the lattice which are sharply peaked on isotropic, flat cosmology several simplifications are possible which reduce the computational effort. We present a step-by-step algorithm resulting in an analytical expression including up to first order corrections in the spread of the state. The algorithm is developed in such a way that it makes the computation straightforward and easy to be implementable in programming languages such as Mathematica.\\
Exemplarily, we demonstrate how the algorithm streamlines the road to obtain the expectation value of the euclidean part of the scalar constraint and as a consistency check perform the analytic computation as well. To showcase further applications of the algorithm, we investigate the fate of the effective dynamics program custom in Loop Cosmology and find that the next-to-leading order corrections can {\it not} be used as corrections for an effective Hamiltonian.
}
\end{abstract}

\maketitle
\section{Introduction}
\label{s1}
\numberwithin{equation}{section}
With the beginning of the era of gravitational wave astronomy and the possibility at reach to probe physics beyond the CMB, more interest is put in recent years on the potential first effects a theory of quantum gravity can have on isotropic cosmological models. One candidate for quantum gravity comes in the form of Loop Quantum Gravity (LQG) \cite{GP00,Rov04,Thi09} whose development has pushed consistently towards computability and phenomenology in the past decade. The construction of the theory is not fully finished yet, in the sense that there remain quantisation ambiguities of the scalar constraint operator, which drives the dynamics of LQG. But while current statements about the dynamical predictions of some incarnation of the quantum constraint have to be taken with a grain of salt, renormalisation programms have entered LQG \cite{Oek02,Bahr:2015,BahrDittrich09,LLT,Thi20} in order to address this caveat. Simultaneously to this development, however, one must ask how to proceed towards observations once a reliable quantum constraint is obtained.\\
Isotropic cosmologies of Friedmann-Lema\^{i}tre-Robertson-Walker (FLRW) type  \cite{Fr22,Le31,Rob35,Wal37} provide the most promising starting point as the overall large scales of the universe today as well as in an epoch before formation of the CMB seems to support this model. With the CMB being mostly classical, quantum corrections to the universe must be small and can only be more dominant in the far past, thus suggesting to use the description of semi-classical, coherent states, whose quantum evolution behaves classical into the future and potentially leads to a resolution of the singularity in the far past. While the construction of such states can not be completed until the above-mentioned quantisation ambiguities are fixed, good candidate coherent states, so called Gauge field theory Coherent States (GCS), have already been developed in LQG by Thiemann and Winkler\cite{Hall94,Hall97,TW1,TW2,TW3}. The idea would be to use these states peaked on isotropic cosmology and supported on a finite lattice, to get a first intuition on the quantum effects of general relativity. Since in the limit of vanishing spread of the states, the classical theory on the lattice is recovered (for large momenta), one has to consider a finite spread $t$ in order to get a first intuition on the quantum effects. Thus, one must be interested in a small value of $t$, whose linear corrections to the classical order are not negligible. However, as soon as one considers a lattice, i.e. many countable degrees of freedom, the problem  typically becomes very involved. The first investigations in this direction \cite{QRLG1} constructed the isotropic cosmological coherent state as a tensor product of Thiemann-Winkler GCS on each edge of the lattice. However, they could only deal with the lattice content when truncating the theory and transcended to the toy model of Quantum Reduced Loop Gravity \cite{QRLG2,Mak20}. However, with such simplification, knowledge of the real quantum effects is lost and it thus remained an open issue how to proceed. Lattice Gauge Theory (to which LQG is closely related) knows the issue of involved lattice computations and relies mostly on numerical methods to tackle it \cite{KS75,Cre83}. In LQG, by now understanding of its kinematical objects has developed so far that we can give a clear guidance on how to approach such computations analytically.\\
The crucial starting point is the set-up of Algebraic Quantum Gravity (AQG) \cite{AQG1} by Giesel and Thiemann, where LQG is restricted to a cubic lattice. In \cite{AQG2}, the authors managed to replace inside of coherent state expectation values the paramount Ashtekar-Lewandowski volume operator with another operator, which is polynomial in the basic configuration variables with arbitrary error-control. We refer to this replacement as the Giesel-Thiemann volume as it sparked the investigation of isotropic coherent states of full Loop Quantum Gravity on the lattice. The first computations were performed in \cite{DL17b}, proving several important statements about the expectation values of flux-polynomial operators of GCS on single edges, which decreased the {\it computational effort by an exponential factor}, when interested in corrections including the first order in the spread $t$ of the states. However, it was not until this year, when \cite{LZ20} delivered a {\it general formula to compute the expectation value of arbitrary polynomials}, i.e. including any powers of holonomies, up to first order in $t$. Both of these latter works paved the way to develop the streamlined algorithm of this paper: we give a step-by-step guide on how to compute the expectation values of any polynomial operator in isotropic coherent states of full Loop Quantum Gravity on a lattice.\\
%Moreover, this 'computer code' has been implemented in the programming language Mathematica allowing the straightforward extraction of these operators. This paper deals mostly with the algorithm to make it available to the community and we shift applications of it to later publications, after highlighting an important consequence of the linear corrections to the effective dynamics program commonly employed in Loop Quantum Cosmology.\\

This article is organized as follows. In section \ref{s2}, we repeat the construction of cosmological coherent states in LQG (in the appendix we also state the necessary background of LQG). The basis elements of the kinematical Hilbert space are supported on finite graphs. Therefore, gauge coherent states can be used to describe a semi-classical geometry on each graph. Following \cite{DL17b}, this is explicitly done for a cubic lattice and isotropic, flat cosmology as underlying geometry. Importantly, we recall the expectation value of polynomial operators on a single edge each up to first order in the spread from \cite{LZ20}. Finally, we recall and introduce new material to facilitate the developments presented in the following section.

In section \ref{s3}, we construct a step-by-step algorithm, which provides a hand-on tool to compute the expectation value of any polynomial operator in the basic variables (holonomies and fluxes) in the cosmological coherent states of the previous section. Two crucial points simplify the investigation: (i) the realisation from \cite{DL17b} that the right-invariant vector fields only give a leading order contribution when their index equals zero, which removes all but few combinatorial cases to consider and decreases the computational time by an exponential factor and (ii) a strategy to extract the presence of $\hat{Q}_v$ operators (the building block of the volume operators) out of the expectation value, reducing the number of right-invariant vector fields on each edge, again speeding up the computation.\\
In section \ref{s4}, we test our algorithm on $\hat{C}_E^\epsilon$ the Euclidean part of the scalar constraint in its common quantisation by Thiemann \cite{Thi96_1} to highlight how much the computation simplifies.

In section \ref{s_effectivedynm}, we further use the algorithm to extract knowledge about the commutator between $\hat{C}^\epsilon_E$ and the volume operator. We find that the real quantum corrections do not match with what one would expect from the {\it effective dynamics program}: here, one typically interprets the expectation value as a function on the classical phase space and evolves classical observable with respect to it. However, when finite $t$ corrections at first order are present, the two procedures `taking the expectation value' and `applying the commutator/Poisson bracket' do not commute anymore. Thus, we find the effective dynamics program to be {\it invalidated} for lattice Loop Quantum Gravity.

In section \ref{s_concl} we finish with conclusion and outlook for the applications of the algorithm as next steps in the cosmological coherent state expectation value endeavour.\\

\section{Cosmological Coherent States}
\label{s2}
\numberwithin{equation}{section}
In this section, we revise the necessary concepts of gauge field coherent states and their application to isotropic, flat cosmology. The structure follows mostly \cite{DL17b}. For further details we refer to the literature. The experienced reader may jump directly to section \ref{s3}.

\subsection{Recap of the Gauge Field Coherent States (GCS) in LQG}
In order to clarify notation, this section will summarize the main concepts from Gauge Field Theory Coherent States (GCS) for canonical Loop Quantum Gravity. For further details we refer to the appendix or the literatur, e.g. \cite{Thi09}.\\

Loop Quantisation of General Relativity promotes the phase space over a graph $\gamma$ from (\ref{PB_HolHol})-(\ref{PB_FluFlu}) to a quantum Hilbert space $\Hil_\gamma$ with suitable operators thereon. The union of {\it all} single graph Hilbert spaces describes then the Ashtekar-Lewandowski Hilbert space $\Hil_{AL}$, which is the kinematical Hilbert space of LQG. For the purpose of this article, however we will later on restrict our attention to a single graph $\gamma$ and study only states in $\Hil_{AL}$ which have excitations thereon.\\
Given a graph $\gamma$, one assigns to each edge $e$ a function in $\Hil_e=L_2({\rm SU}(2),d\mu_H)$ with $\mu_H$ being the unique left- and right-invariant Haar measure over ${\rm SU}(2)$. The  full Hilbert space $\Hil_\gamma$ is then simply the tensor product over all square integrable functions on each edge, $\Hil_\gamma:=\otimes_{e} \Hil_e$. The holonomies get promoted to bounded, unitary multiplication operators and the fluxes to essentially self-adjoint derivation operators. If we label in the position representation $\psi\in\Hil_\gamma$ as $\psi=\psi(\{g_e\}_{e\in\gamma})$ then:
\bal \label{quantisation}
\hat{h}^{(k)}_{mn}(e')\psi(\{g_e\}_{e\in\gamma})&:=D^{(k)}_{mn}(g_{e'})\psi(\{g_e\}_{e\in\gamma}),\\
\hat{P}^K(e')\psi(\{g_e\}_{e\in\gamma})&:=-\frac{i\hbar\kappa\beta}{2}R^K(e')\psi(\{g_e\}_{e\in\gamma}),
\eal
with the  $D^{(k)}_{mn}(g_e)$ being the Wigner-matrix of group element $g_e$ in the $2k+1$-dimensional, irreducible representation of ${\rm SU}(2)$ and the {\it right-invariant vector field} $R^K(e)$:
\bal\label{right_inv_vf}
R^K(e')\psi(\{g_e\}_{e\in\gamma}):=\left.\frac{d}{ds}\right|_{s=0}\psi(...,e^{s\tau_K}g_{e'},...),
\eal
which obey the following commutation relation with $f^{IJ}_K$ denoting the structure functions of $\mathfrak{su}(2)$ with respect to the used basis:
\bal\label{CommRel}
&[R^K(e),\hat{h}^{(j)}(e')_{ab}]=\delta_{e'e}[\tau^K]^{(j)}_{ac}\hat{h}_{cb}(e),\hspace{30pt} [R^I(e),R^{J}(e')]=-i\delta_{e'e}f^{IJ}_{K}\; R^K(e).
\eal
For the definition of $[\uptau^K]^{(j)}$ see appendix and equation (\ref{structure_function_tau}).
In addition, we introduce the momentum operator acting on the other end of the edge
\begin{align}
\hat{P}^K({e'}^{-1})\psi(\{g_e\}_{e\in\gamma})&:=\frac{i\hbar\kappa\beta}{2}L^K(e')\psi(\{g_e\}_{e\in\gamma}),
\end{align}
with left-invariant vector fields $L^K$ defined analogously to (\ref{right_inv_vf}) and obeying similar commutation relations \cite{DL17b}.\\
The $\tau_I$ label a suitable basis of $\mathfrak{su}(2)$. We will work mostly with the two following choices:
For the Pauli matrices $\sigma_I$ we denote $\underline{\tau}_I:=-i\sigma_I/2$ with $I=1,2,3$, for which $[\underline{\tau}_I,\underline{\tau}_J]=\underline{\epsilon}_{IJK}\underline{\tau}_K$ with $\underline{\epsilon}_{IJK}$ being the Levi Civita symbol. Alternatively, the {\it spherical basis} is defined as follows for $S\in\{-1,0,+1\}$, where $\uptau_{\pm}:=\pm (\tau_1\pm i\tau_2)/\sqrt{2}$ and $\uptau_0:=\underline{\tau}_3$ and which are subject to the algebraic relations $[\uptau_+,\uptau_-]=i \uptau_0$, $[\uptau_{\pm},\uptau_0]=\pm i\uptau_{\pm}$. If the basis is unspecified we will use the notation $\tau_I$ and $\epsilon_{IJK}$.\\

We will now describe the form and properties of the Gauge Field Theory Coherent States. For their derivation, the reader is referred to the literature \cite{TW1,TW2,TW3}.\\
Based on the idea of approximating a given classical field configuration $(\tilde{A},\tilde{E})$, one computes first for each lattice edge $e$ the corresponding smeared quantities $(\tilde{h}(e), \tilde{P}(S_e))$ and maps this to a complex polarisation of the classical phase space, i.e. $(\tilde{h}(e), \tilde{P}(S_e))\mapsto h^\mathbb{C}_e\in {\rm SL}(2,\mathbb{C})$ that expresses the complex connection as a function of the real phase space. For example, the left-polar decomposition prescribes:
\bal
h^{\mathbb{C}}_e:=\exp\left(-\frac{it}{\hbar \kappa} \tau_J\tilde{P}^J(S_e)\right)\tilde{h}(e).
\eal
The so called {\it semiclassicality parameter} $t:=\hbar \kappa/\ell^2 \geq 0$ is dimensionless where $\ell$ is some length scale.
\begin{Definition}[GCS]
The \emph{Gauge Field Theory Coherent State} $\psi_{h^\mathbb{C}}^t\in\mathcal{H}_e$ for each edge $e$ and classical field configuration $h^{\mathbb{C}} \in {\rm SL}(2,\mathbb{C})$ is given by 
\bal\label{lpDecomp}
\psi^t_{h^\mathbb{C}}(g):=\sum_{j\in\mathbb{N}_0/2}d_j e^{-t(d_j^2-1)/8} {\rm Tr}^{(j)}\left( h^\mathbb{C} g^\dagger \right),
\eal
where $t\geq 0$, $d_j=2j+1$ and ${\rm Tr}^{(j)}(.)$ denotes the trace in the spin-$j$ irreducible representation of $SU(2)$. Finally, $\langle1\rangle:=||\psi^t_{h^\mathbb{C}}||^2$ denotes the normalisation of the state.
\end{Definition}

As was shown in \cite{TW2} these GCS are sharply peaked on the classical configuration in the following sense:
\begin{Theorem}
Let $\psi^t_{h^\mathbb{C}},\psi^t_{g^\mathbb{C}}\in\mathcal{H}_e$ be two GCS. For all $h^\mathbb{C},g^\mathbb{C}\in {\rm SL}(2,\mathbb{C})$ there exists a positive function $K_t(h^\mathbb{C},g^\mathbb{C})$ decaying exponentially fast as $t\rightarrow0$ for $h^\mathbb{C}\neq g^\mathbb{C}$ and such that
\bal
|\langle\psi^t_{h^\mathbb{C}},\psi^t_{g^\mathbb{C}}\rangle|^2\leq K_t(h^\mathbb{C},g^\mathbb{C})||\psi^t_{h^\mathbb{C}}||^2 ||\psi^t_{g^\mathbb{C}}||^2.
\eal
Moreover, for holonomy and flux operators on $\mathcal{H}_e$ one finds 
\bal\label{CohStaDef}
\langle \psi^t_{h^\mathbb{C}},\hat{h}_{ab}(e)\psi^t_{g^\mathbb{C}}\rangle&=h_{ab}(e)\langle\psi^t_{h^\mathbb{C}},\psi^t_{g^\mathbb{C}}\rangle+\mathcal{O}(t),\\
\langle \psi^t_{h^\mathbb{C}},\hat{P}^J(S_e)\psi^t_{g^\mathbb{C}}\rangle&=P^J(S_e)\langle \psi^t_{h^\mathbb{C}},\psi^t_{g^\mathbb{C}}\rangle+\mathcal{O}(t),
\eal
where $h(e)$ and $P^J(e)$ stem from the decomposition of $h$ in (\ref{lpDecomp}).
\end{Theorem}

We will end this subsection with recalling an important statement regarding the volume operator of LQG. The quantisation of the volume as the Ashtekar-Lewandowski volume operator $\hat V_{AL}$ \cite{AL97_Vol} proved vital for many constructions of other important operators such as the quantisation scalar constraint (see \cite{Thi96_1,Thi96_2} and later in section \ref{s4}). Albeit it being a highly complicated object whose spectrum is not yet under full control, computations of physical relevant quantities are possible due to the following crucial observation from \cite{AQG2}:

\begin{Theorem} 
For any gauge field theory coherent state $\Psi_\gamma^t$ defined on a graph $\gamma$, the Ashtekar-Lewandowski volume operator at a vertex $v\in \gamma$, defined as
\begin{align}
\hat V^{AL} _v &= \frac{(\beta\hbar\kappa)^{3/2}}{2^{7/2}\sqrt{3}}\sqrt{|\hat Q_v|},\\
\hat Q_v &:= -i(\frac{2}{\hbar\kappa\beta})^{-3} \sum_{ijk}\epsilon(i,j,k)\epsilon_{IJK} \hat P^I(e_i)\hat P^J(e_j)\hat P^K(e_k),
\end{align}
with $\epsilon(i,j,k):=sgn(\det(\dot{e}_i,\dot{e}_j,\dot{e}_k))$, and the $k$-th Giesel-Thiemann volume operator, defined as
\begin{align}\label{giesel-thiemann-volume}
\hat{V}^{GT}_{k,v}=\frac{(\beta\hbar\kappa)^{3/2}}{2^{7/2}\sqrt{3}}\sqrt{\langle \hat Q_v\rangle}\left[\id_{\Hil_\gamma}+\sum_{n=1}^{2k+1}\frac{(-1)^n}{n!}\left(n-1-\frac{1}{4}\right)!\left(\frac{\hat Q_v^2}{\langle \hat Q_v\rangle^2}-\id_{\Hil_\gamma}\right)^n\right],
\end{align}
with the notation $\langle \hat Q_v\rangle:=\langle \Psi^t_\gamma, \hat{Q}_v \Psi^t_v\rangle$, have the same expectation value for any polynomial $F$ involving either of them up to order $t^{k}$, i.e.:
\begin{align}
\langle \Psi^t_\gamma,F(\hat V^{AL}_v,\hat{h})\Psi^t_\gamma\rangle=\langle \Psi^t_\Gamma, F(\hat{V}^{GT}_{k,v},\hat{h})\Psi^t_\gamma\rangle+\mathcal{O}(t^{k+1}).
\end{align}
\end{Theorem}

Let us also state a slight reformulation of $\hat{V}^{GT}_{k,v}$ which is used in the next section:
\bal\label{volexp}
\hat{V}^{GT}_{k,v}=\frac{(\beta \hbar \kappa)^{3/2}}{2^{7/2}\sqrt{3t}}\sum_{N=0}^{4k+2} c_N \, \hat{Q}_v^N,
\eal
where the coefficients $c_N$ are determined via the series in (\ref{giesel-thiemann-volume}).

\subsection{Recap of GCS in isotropic, flat cosmology on a fixed graph}
We apply the theoretical framework of the last two subsections to isotropic, flat cosmology, also known as Friedmann-Lema\^itre-Robertson-Walker (FLRW) spacetime.\\

Typically, flat FLRW is studied in the context of the $\sigma=\mathbb{R}^3$ as spatial manifold. However, for the moment we restrict our attention to the compact torus $\sigma_{\cal R}=[0,\mathcal R]^3$, with infra-red cut-off ${\cal R}\in\mathbb{R}$. After all quantum computations have been evaluated in presence of this finite cut-off, it will be trivial to perform the thermodynamical limit $\cal R\to \infty$.\\
%The Robertson-Walker metric is parametrised in the following way: ($a,b,...=1,2,3$)
%\bal
%ds^2=-N^2dt^2+a(t)^2(dx^2+dy^2+dz^2),
%\eal
%in a suitable gauge-fixing of the diffeomorphism constraint in which explicitly the shift vector vanishes and the spatial metric reads $q_{ab}(x)=a(t)^2 \delta_{ab}$ with the {\it scale factor} $a(t)$.\\
In isotropic, flat cosmology, the Ashtekar-Barbero variables take the form
\bal\label{cos_con_tri}
A^J_a=c\; \delta^J_a,\hspace{30pt}E^a_J=\bar{p}\; \delta^a_J.
\eal
We will now introduce a discretisation of $\sigma_{\cal R}$ in form of  a cubic lattice $\gamma$ with $M$ points in each direction of the coordinates described by $x^a$. With respect to the fiducial flat metric the coordinate length of the torus was $\cal R$ and we will denote by $\epsilon={\cal R}/M$ the regulator of the discretisation, i.e. the coordinate length of each edge. We will compute the holonomies $h(e)\in {\rm SU}(2)$ of the connection, which for an edge $e_k$ along direction $k\in \{\pm 1,\pm 2,\pm 3\}$ read:
\bal
h(e_k)=\mathcal{P}\exp\left(\int_0^1dt\; A^J_a(e_k(t))\,\tau_J\,\dot{e}^a(t)\right)=\exp(sgn(k)\,c\,\epsilon\, \tau_{|k|}).
\eal
Note that $D^{(j)}_{mn}(e^{z\uptau_0})=\delta_{mn}e^{-izm}$ and thus $U^{1/2}(k)\equiv h(e_k)$ with the definition:
\bal\label{classical_hol}
U^{j}(k):=D^{(j)}(h(e_k))=D^{(j)}(e^{\xi \tau_k})=D^{(j)}(e^{\xi \vec{n}_{k}\cdot\vec{\tau}})=D^{(j)}(n_k e^{\xi \uptau_0}n_k^\dagger),
\eal
where we introduced the SU(2) rotations:
\begin{align}\label{n_matrix_definition}
    n_1 = \frac{1}{\sqrt{2}}\left(\begin{array}{cc}
    1     & -1 \\
        1 & 1
    \end{array}\right),\hspace{20pt}    n_2 = \frac{1}{\sqrt{2}}\left(\begin{array}{cc}
    1     & i \\
        i & 1
    \end{array}\right),\hspace{20pt}    n_3= \left(\begin{array}{cc}
    1     & 0 \\
        0& 1
    \end{array}\right).
\end{align}

For the gauge-covariant fluxes \cite{Thi00_gc}, we have to choose a certain set of paths $\rho_x$ in their construction (\ref{gaugecov_fluxes}). For an edge $e_k$ one splits $\rho_x=\rho_{x,a}\circ \rho'_{x,b}$ where $\epsilon_{kab}=1$. We choose further that $\rho_{x,a}[0]=e_k\cap S_{e_k}$ and $\rho_{x,a}[1]=\rho'_{x,b}[0]$ and $\rho'_{x,b}[1]=x$. Finally, the path $\rho_{x,a}$ starts in direction $\pm a$, stays in $S_{e_k}$ with tangent constant vector, and similar for $\rho'_{x,b}$ with constant tangent vector oriented along $\pm b$. For the choice, the gauge-covariant fluxes for connection and triad (\ref{cos_con_tri}) is found to be:
\bal\label{gauge_cov_p}
P^I(e_k)= \delta^I_k\, p\, :=  \delta^I_k\,\bar{p}\,{\rm sinc}(c\,\epsilon/2)^2.
\eal
%We want to stress again, that this is a departure from the standard fluxes used earlier in \cite{DL17b} and other constructions in the literature (see e.g. \cite{QRLG1}). However, passing to these gauge-covariant fluxes is necessary in order to guarantee the existence of discretisations of the geometrical constraints that are intrinsically invariant under ${\rm SU}(2)$-rotations.\\
We will construct our cosmological coherent state as follows: for each edge, going in the same direction $I$, we peak on the same element $H_I\in $SL(2,$\mathbb{C}$), thus capturing homogeneity. Each element looks as follows in its holomorphic decomposition: \cite{Car00,BMP09}
\begin{align}
    H_I = n_I\, e^{-(\xi-i\eta)\uptau_0}n_I^\dagger,\hspace{40pt}\eta =\frac{2\epsilon^2\,p}{\ell^2\beta},\hspace{20pt}\xi=\epsilon\,c,
\end{align}
with $n_I$ from (\ref{n_matrix_definition}), $c$ from (\ref{cos_con_tri}) and $p$ defined in (\ref{gauge_cov_p}).\footnote{Note, that we have not peaked on $\bar{p}$, but on $p$ which includes the gauge-covariant corrections. This is congruent with general relativity on a graph restricted to cosmology, see \cite{wojtek} for further details.} 
The {\it isotropic, flat cosmology coherent state} $\Psi_\gamma^t$ on a fixed cuboidal lattice $\gamma$ of $M$ many vertices is  given by the tensor product of the GCS $\psi^t_{H_I}$ on each edge:
\begin{align}\label{cosmological_coherent_state}
    \Psi^t_\gamma (\{g\}):= \prod_{I\in\{1,2,3\}}\prod_{k\in\mathbb{Z}^3_M} \psi_{H_I}^t(g_{k,I}).
\end{align}
We finish this section with the important statement, how the expectation values of arbitrary polynomial operators on a single edge in $\psi^t_{H_I}$ look including up to next-to-leading order. The proofs of the formulas provided below are shown in \cite{DL17b} and \cite{LZ20} for $\eta\neq 0$:
\bal\label{normalisation}
\langle1\rangle:&=\langle\psi^t_{H_I}, \hat{h}^{(0)}_{00} \psi^t_{H_I}\rangle=
\sqrt{\frac{\pi}{t^3}}\frac{2\eta \,e^{\eta^2/t}}{\sinh(\eta)}e^{t/4},
\eal
is the normalisation of each state. The main result is expressed using the spherical basis, i.e. $K_1,...,K_N\in\{-1,0,+1\}$ and $\hat h\equiv\hat h (e)$, i.e. going in the same direction as the edge:
\bal
\langle\psi^t_{H_I},\hat{h}^{(k)}_{ab}R^{K_1}...R^{K_N}\psi^t_{H_I}\rangle &=
\langle\psi^t_{H_I}, \hat{h}^{(k)}_{cb}\psi^t_{H_I}\rangle \left(\frac{i\eta}{t}\right)^N  D^{(1)}_{-K_1-S_1}(n_I)...D^{(1)}_{-K_N-S_N}(n_I)\; \Bigg(\delta^{S_1}_0...\delta^{S_N}_0\delta_{ac}+\nonumber\\
&\hspace{10pt}+\frac{t}{2\eta}\bigg[\delta_{ac}\delta^{S_1}_0...\delta^{S_N}_0
N\left(\frac{N+1}{2\eta}-{\rm coth}(\eta)\right)+\label{hRRR}\\
&\hspace{10pt}+i\sum_{A=1}^N \delta^{S_1}_0...\cancel \delta^{S_A}_0 ...\delta^{S_N}_0  (1-s_A {\rm tanh}(\eta/2))D^{(k)}_{-s_A-L}(n^\dagger_I)[\uptau^{L}]^{(k)}_{ac}\nonumber\\
&\hspace{10pt}\left.-\frac{\delta_{ac}}{\sinh(\eta)}\sum_{A<B=1}^N\delta^{S_1}_0...\cancel \delta^{S_A}_0...\cancel \delta^{S_B}_0... \delta^{S_N}_0(\delta^{S_A}_{+1}\delta^{S_B}_{-1}+\delta^{S_A}_{-1}\delta^{S_B}_{+1})e^{S_A\eta}\bigg]
\right)\nonumber+\mathcal{O}(t^2),
\eal
and
\bal\label{Result}
\langle\psi^t_{H_I}, \hat{h}^{(j)}_{ab} \psi^t_{H_I}\rangle = \langle1\rangle \; \sum_{c=-j}^j D^{(j)}_{ac}(n_I) e^{-i\xi c} \gamma^j_c D^{(j)}_{cb}(n^\dagger_I),
\eal
with $\gamma^j_c=1-t \tilde\gamma^j_c$ where
\bal\label{GammaResult}
\tilde\gamma^j_c=\frac{1}{4}\left[(j^2+j-c^2)\frac{{\rm tanh}(\eta/2)}{\eta/2}+c^2\right]+{\cal O} (t^2).
\eal

\subsection{Useful properties for cosmological coherent state expectation values}
Before presenting a step-by-step guide in the next section, we will use this section to recall some already known properties as well as some new ones, which simplify the future computations. And by explaining the origin of the induced simplifications here, we keep the algorithm itself slim.\\
We use the spherical basis $K\in\{-1,0,+1\}$ for the right-invariant vector fields throughout this section.\\

\begin{obser}\label{obs1}  In \cite{DL17b} it was observed that only the leading order contribution comes from the first line of  (\ref{hRRR}) which is $\delta^{S_1}_0...\delta^{S_N}_0 \delta_{ab}$ that is, for the leading order contribution, one does not have to perform any contractions. Likewise when investigating the next lines, one sees that the first order corrections come from the contractions of at most a single or one pair of indices being non vanishing.
\end{obser}

\begin{obser}\label{obs2}
Another observation from \cite{DL17b} is that the total numbers of right/left-invariant vector fields determines the leading order: each flux being proportional to $t^{-1}$ at most: A monomial with $N$ many right/left-invariant vector fields has an expectation value of order $\mathcal{O}(t^{-N})$.
Thus, in a sum together with other polynomial operators with a $N-2$ or fewer fluxes they can be neglected due to being higher order corrections than we are interested in.\\
Note, that at next-to-leading order operators may be commuted freely, as due to (\ref{CommRel}) each commutation removes a right-invariant vector field.
\end{obser}

\begin{obser}
Again from \cite{DL17b} we utilize the knowledge that any of the three directions can be brought into the form of computing only the expectation value on $\psi^t_{H_3}$ by extracting the corresponding $n_i$ matrices. Further, for a $\hat F$ a polynomial operator we can replace  the {\it right-most} left-/right-invariant vector fields:
\begin{align}\label{replaceLtoR}
    \langle\psi^t_{H_3}, \hat F \; L^K \psi^t_{H_3} \rangle = e^{-i\,z\,K} \langle \psi^t_{H_3}, \hat F \; R^K \psi^t_{H_3} \rangle,
\end{align}
with $z=\xi -i \eta$. Further, one can show that
\begin{align}
    [R^K,L^J]=0.
\end{align}
Hence, when having a string of left-invariant vector fields after replacing the right-most one it can be freely brought to the left end of the string.
\end{obser}

\begin{obser} In \cite{DL17b}, it was first observed that for $a=1,2,3$:
\begin{align}
D^{(1/2)}(n_a)\uptau_0D^{(1/2)}(n_a^\dagger)&=\underline{\tau}_a,
\end{align}
and
\begin{align}
\epsilon^{(n:\mu_1\mu_2\mu_3)}_{K00}=\epsilon^{(n:\mu_1\mu_2\mu_3)}_{0K0}=\epsilon^{(n:\mu_1\mu_2\mu_3)}_{00K}&=-\delta_{0K}\label{EpsilonTrick},
\end{align}
where ( $I,J,K,A,B,C\in\{-1,0,+1\}$)
\bal
 \epsilon^{(n:\mu_1\mu_2\mu_3)}_{ABC} := \epsilon^{IJK}D_{-I-}(n_{|\mu_1|})D_{-J-B}(n_{|\mu_2|})D_{-K-C}(n_{|\mu_3|}),\hspace{20pt}
\epsilon^{IJK}:=i\sqrt{6}\left(\begin{array}{ccc}
1 & 1 & 1\\
I & J & K
\end{array}\right),
\eal
which is  anti-cyclic in its indices in the following sense:
\begin{align}
    \epsilon^{(n:\mu_1\mu_2\mu_3)}_{ABC}=-\epsilon^{(n:\mu_2\mu_1\mu_3)}_{BAC}.
\end{align}
\end{obser}

The above tools lead eventually to the expectation value of $\hat{Q}$ in cosmological coherent states
\begin{align}
q_o:=\frac{ 48 \eta^3}{t^3}\left(1+t\frac{3}{2\eta^2}(1-\eta \, {\rm coth}(\eta)\right)=\langle\Psi^t_\gamma,\hat{Q}_v\,\Psi^t_\gamma \rangle+\mathcal{O}(t^{-1}).
\end{align}

\begin{obser} \label{obs5}
Let $\hat{F}=\prod_e l_e$ be an operator where each $l_e$ is a monomial supported on the edge "$e$" of the form $\hat{h}^{(j)}_{ab} R^{K_1}... R^{K_{N^e}}$. We introduce a \emph{modified average}:
\begin{align}
    \langle\langle l_e \rangle\rangle := \langle\psi_{H(e)}^t,l_e \;\psi^t_{H(e)}\rangle+ \delta^{K_1\ldots K_{N^e}}_0\;  Y^{(j)}_{ab}[v,m,N],
    \end{align}
with:
\begin{equation}
Y^{(j)}_{ab}[v,m,N] = \frac{i}{4}\left(\frac{i \eta}{t}\right)^{N^e-1} \left(2N^e+ N_v+N_{v+e_m}\right)  \sum_{c=-j}^j c\, e^{-i\xi c} D^{(j)}_{ac}(n_m) D^{(j)}_{cb}(n_m^\dagger).
\end{equation}
We further define
\begin{align}
     \tilde l_e =t^{-N^e}\lim_{t\rightarrow0}\left(t^{N^e} \langle\langle l_e \rangle\rangle \right),
\end{align}
being the leading order of $l_e$, i.e., $ \langle\langle l_e \rangle\rangle/\tilde l_e-1 =\mathcal{O}(t)$. Then, the following statement holds at the classical order:
\begin{align}\label{Q_trick_classical}
    \langle \Psi^t_\gamma, \hat {F} \prod_v \hat Q_v^{N_v}\Psi^t_\gamma\rangle =P_o+\mathcal{O}(t),\hspace{30pt}P_o:= \left(\prod_v 2^{3N_v}\left(\frac{\eta i}{t}\right)^{3N_v} (-6i)^{N_v}\right) \prod_e \tilde{l}_e,
\end{align}
and first order corrections can be obtained from
\begin{align}\label{Q_trick_quantum}
    \langle \Psi^t_\gamma,\hat F\; \prod_v \hat Q_v^{N_v}\Psi^t_\gamma \rangle  =& P_o \bigg(
        1+\sum_e \frac{1}{\tilde l_e}    \big[ \langle\langle l_e \rangle\rangle - \tilde l_e +
        \\
        &+
        \frac{t}{16\eta^2} (1-\delta_{N_v,0}\delta_{N_{v+e},0})(N_v+N_{v+e})(4N^e+N_v+N_{v+e}+3-4\eta \coth(\eta))
        \tilde l_e
        \big]
    \bigg)
    +\mathcal{O}(t^2).\nonumber
\end{align}
\end{obser}

The proof of this statement can be obtained by utilizing (\ref{EpsilonTrick}) and after several pages of computation and is  therefore omitted at this point. Importantly, this property allows to get rid of all appearances of the $\hat Q_v$ {\it at one instant}. Thus, one no longer needs to sum over all the indices but one only has to add the corresponding quantum correction at the end.

\section{Algorithmic approach to expectation values with cosmological coherent states}
\label{s3}
\numberwithin{equation}{section}

On a formal ground, the cosmological coherent states $\psi^t_\gamma$ can serve as a powerful mean to describe the semi-classical behaviour of quantum cosmology studied from the perspective of the full theory of Loop Quantum gravity put on a lattice. However, one can quickly face obstacles of a purely practical origin, namely, the expressions to be calculated are intractable, or at least seem to be such. As a step towards the solution to this problem we construct \textit{an algorithm} which allows for a simplification of generically complex expressions, having in mind that at the moment we are only interested in the first-order quantum corrections. 

In a nutshell, we classify terms to be further processed (evaluated) as \textit{classical} or as (first order) \textit{quantum}, depending on the order of $t$ they carry. It turns out, that in our approach we split the average value of an operator of interest into a moderate number of classical contributions, which can in principle be cumbersome, and a typically huge number of quantum corrections, which are much simpler to handle. Note that the classical contributions do not solely consist of terms of a lower order with respect to $t$, in comparison with the quantum ones. They are rather sums of both orders, to be further processed. After we introduce a few pieces of notation necessary for this section we will come back to a formal description of this splitting as well as the whole approach.

\subsection{Discretization}

From now on, we specialize $\gamma$ to be a cubic lattice with $M$ vertices, i.e. each edge $e\in\gamma$ can be parametrized in the following way:
\begin{align}
    e=(v,\mu),\hspace{30pt} \mu\in \{1,2,3\}, \;\;v \in\mathbb{Z}^3_{M}=\{0,...,M-1\}^3.
\end{align}
We introduce an auxiliary notation by which each edge gets associated with a second label, namely
\begin{align}
    e_{v,\mu}=(v,-\mu), \hspace{30pt} \mu \in \{1,2,3\}.
\end{align}
Of course, both notations can be easily translated into each other by identifying
\begin{align}
    (v,\mu) = (v+e_\mu, -\mu)
\end{align}
for which we further defined the unit vectors $e_\mu$ whose $I$th components $e_\mu^I$ read
\begin{align}
e_\mu^I = \text{sgn}(\mu) \delta^I_{|\mu|}.
\end{align}
This parametrization allows us to write some ingredients more explicitly, e.g.
\begin{align}
    \epsilon(i,j,k):=sgn(\det(\dot{e}_i,\dot{e}_j,\dot{e}_k))={\rm sgn}(\mu_i\mu_j\mu_k)\epsilon_{|\mu_i||\mu_j||\mu_k|},
\end{align}
With the help of these vectors we later employ periodic boundary conditions. To this end we will make the identification of the vertices $M e_\mu \rightarrow 0\; e_\mu$, that is, $v(M)=v(0)$.

Most importantly, the action of the operators involved becomes:
\begin{align}
    \hat P^K(v,\mu)&=\mathcal E\begin{cases}
R^K(v,\mu) & if\hspace{5pt}\mu>0\\
-L^K(v+e_{\mu},-\mu) &if\hspace{5pt}\mu<0
\end{cases},\label{fluxes}
\\
\hat h_{ab}^{(j)}(v,\mu)&=\begin{cases}
D_{ab}^{(j)}(v,\mu) & if\hspace{5pt}\mu>0\\
\left[D^{(j)}_{ab}(v+e_{\mu},-\mu)\right]^\dagger &if\hspace{5pt}\mu<0
\end{cases},\label{holonomies}
\end{align}
where $\mathcal E =|i\hbar\kappa\beta/2| \equiv \hbar\kappa\beta/2$. We took the absolute value because we assume dependence of $\hat P$ only inside the volume operator.

\subsection{Input}
In order to develop the algorithm we first need to specify what, on the conceptual level, can serve the role of its main input, as well as which supplementary information needs to be stored for the sake of practical implementations of the algorithm. 

Concerning the first, fundamental question, we assume that the operator for which we will compute the expectation value can only depend on the holonomies $\hat h_{ab}^{(j)}(v,\mu)$ and the volume $\hat{V}_v^{AL}$. We explicitly exclude dependence on $\hat P^K(v,\mu)$ for the moment, as the most interesting operators of LQG feature this behaviour.% (however it can be easily extended at the cost of a few extract signs).

Elaborating on the supplementary input information,  while implementing the algorithm one carefully needs to collect the indices appearing in the operator in question, which shall be contracted at the end of the computation (important also are appropriate summation ranges of these indices). The latter remark is important as the holonomy is equipped with an index labelling the representation (spin). From the practical perspective it turns out useful (allows to reduce computational time) to predict and fix the maximal representation index which can appear in a given computation.

Last but not least, there is a remaining question of the lattice size $M$. If the algorithm is used just as a road towards fully analytic calculations, there is no need to fix its value. However, in a scenario such as implementation on a cluster we need to fix $M$ in a way which minimizes the time necessary for the computation and at the same time guarantees that the final result will be independent of the lattice size. The condition which facilitates such a choice stems from the periodic boundary conditions. Because of this geometry, $M$ needs to be sufficiently large in comparison to the ``correlation length'' of the evaluated operator. For example, if the operator only involves couplings between the nearest vertices, the choice $M=2$ is sufficient and at the same time optimal.

\subsection{Structure of the algorithm}
The initial step to be performed in the algorithm is an appropriate representation of the volume by means of tractable quantities. In fact, in the assumed approximation (first leading order in $t$) we can set $\hat{V}_v^{AL}\equiv \hat{V}_{k,v}^{GT}$, and replace the latter object by $c_N \hat Q^{N}_v$. Note that $N$ is being added to the list of indices to be contracted at the end of the computation, with its range stemming from Eq. (\ref{volexp}). One also needs to remember about the constant present in (\ref{volexp}).

As a result, we work with an operator which is a polynomial in $\hat h(v,\mu)$ and $\hat Q_v$, likely expressed through commutators of these operators. To go further we create two new objects, $\mathcal P_{cl}$ and $\mathcal P_{qu}$, in which we will collect the aforementioned classical and quantum contributions. At the initial state, we place the whole operator of interest inside $\mathcal P_{cl}$ and we set $\mathcal P_{qu}$ as being empty.

The proper part of our algorithm consists of the following steps:
\begin{itemize}
    \item[I.] To resolve commutators appearing in the evaluated operator, using explicit expressions suitable to keep the track of the $\mathcal{O}(t^1)$ order;
    \item[II.] To use the modified commutation relations from the previous step to shift all the holonomies to the left of every monomial;
    \item[III.] To appropriately handle $\hat P^K(v,\mu)$ operators necessarily appearing due to the commutation relations;
    \item[IV.] To perform a ``link splitting" of every monomial present in both $\mathcal P_{cl}$ and $\mathcal P_{qu}$;
    \item[V.] To simplify the products of the holonomies on the same edge;
    \item[VI.] To employ final replacements of the particular expectation values by specific functions. 
\end{itemize}
Each subsection below provides a more detailed description of the above steps.

\subsubsection{Step I: Resolving the commutators}
In order to simplify the consideration we introduce two new operators
\begin{equation}
\hat E_\pm^K(v,\mu)=\hat P^K(v,\mu) \pm \hat P^K(v,-\mu),\qquad \mu=1,2,3.
\end{equation}
Note that we restricted the range of $\mu$ to its positive values. 
We then search $\mathcal{P}_{cl}$, which contains the operator to be processed, for nested commutators. We take the inner most commutator and replace it according to the following rules (summation convention applies to indices which do not appear in original commutators on the left hand sides):
\begin{subequations}
\begin{equation}
[\hat h^{(j)}_{ab}(v,\mu),\hat h^{(j')}_{cd}(v',\mu')]\rightarrow 0, \quad [\hat Q_v^N,\hat Q_{v'}^{N'}]\rightarrow 0,
\end{equation}
\begin{eqnarray}
[\hat h^{(j)}_{ab}(v,\mu),\hat E^K_\pm(v',\mu')]&\rightarrow& -\mathcal {E} \delta_{v,v'}\left(\delta_{\mu,\mu'}\pm\delta_{\mu,-\mu'}\right)[n\uptau n]^{(\mu,j,K)}_{ac} \hat h_{cb}^{(j)}(v,\mu)\nonumber\\
&+&  \mathcal {E}\delta_{v+e_{\mu},v'}\left(\delta_{\mu,-\mu'}\pm\delta_{\mu,\mu'}\right)\hat h_{ac}^{(j)}(v,\mu)[n\uptau n]^{(\mu,j,K)}_{cb},
\end{eqnarray}
\begin{equation}
    [\hat E^I_\pm(v,\mu),\hat E^J_{\pm'}(v',\mu')]\rightarrow -\mathcal{E}\delta_{v,v'}f_{IJK}\left(\delta_{\mu,\mu'}\pm'\delta_{-\mu,\mu'}\right)\hat E^K_{\pm\cdot\pm'}(v,\mu),
\end{equation}
\begin{equation}\label{Replacement_hQ}
[\hat h^{(j)}_{ab}(v,\mu),\hat Q_{v'}^N]\rightarrow -6 i \mathcal{E}N\mathrm{sgn}(\mu) \left(\hat A^{(j)}_{a b}(v,v',\mu)
\hat Q^{N-1}_{v'}-\Upsilon\frac{N-1}{2}[\hat A^{(j)}_{a b}(v,v',\mu),\hat Q_{v'}]\hat Q^{N-2}_{v'}\right),
\end{equation}
\begin{equation}
[\hat E^P_\pm(v,\mu),\hat Q_{v'}^N]\rightarrow 6 i \mathcal{E}N \delta_{v,v'} \left(\hat B^{P}_{\pm}(v,\mu)
\hat Q^{N-1}_{v}-\Upsilon\frac{N-1}{2}[ \hat B^{P}_{\pm}(v,\mu),\hat Q_{v}]\hat Q^{N-2}_{v}\right),
\end{equation}
where
\begin{equation}
  \hat A^{(j)}_{a b}(v,v',\mu)= \epsilon^{(n:\tilde\mu_1\tilde\mu_2\tilde\mu_0)}_{ILK}\left(\delta_{v,v'}[n\uptau n]^{(\mu,j,K)}_{ac}\hat h^{(j)}_{cb}(v,\mu)
+\delta_{v+e_\mu,v'}\hat h^{(j)}_{ac}(v,\mu)[n\uptau n]^{(\mu,j,K)}_{cb}\right) \hat E^I_-(v',\tilde\mu_1)\hat E^L_-(v',\tilde\mu_2),
\end{equation}
\begin{equation}
  \hat B^{P}_{\pm}(v,\mu)= \epsilon^{(n:\tilde\mu_1\tilde\mu_2\tilde\mu_0)}_{IJK}\hat E^I_-(v,\tilde\mu_1)\hat E^J_-(v,\tilde\mu_2)f_{PKL}\hat E^{L}_\mp(v,\mu),
\end{equation}
\begin{equation}
    [n\uptau n]^{(\mu,j,K)}_{ac}:=
D^{(1/2)}_{aa'}(n_{|\mu|})[\uptau^K]_{a'c'}^{(j)}D^{(1/2)}_{-c-c'}(n_{|\mu|})(-1)^{c-c'}.
\end{equation}
\end{subequations}
In the above formulas we introduced the following additional pieces of notation:
\begin{itemize}
    \item $\tilde\mu_s = \left(|\mu|+s\right)\;\mathrm{mod}_1\;3\equiv \left(|\mu|+s\right)-3 \left\lfloor \left(|\mu|+s\right)-1)/3\right\rfloor $, where $\left\lfloor \cdot\right\rfloor$ is the floor function, defined in terms of the modulo operation with offset $1$. Note that $\tilde\mu_0 = |\mu|$;
    \item The symbol $\pm\cdot\pm'$ is understood as "$+$" if $\pm$ and $\pm'$ are the same signs and "$-$" in the opposite case;
    \item $\Upsilon$ is an additional parameter indicating the splitting between the classical and the quantum part, further used by the algorithm to distribute the terms between $\mathcal{P}_{cl}$ and $\mathcal{P}_{qu}$.
\end{itemize}
After the replacement of the inner most commutators in $\mathcal{P}_{cl}$ is done we update both objects as follows:
\begin{equation}\label{update}
  \mathcal{P}_{qu}:= \mathcal{P}_{qu}+ \left.\frac{d \mathcal{P}_{cl}}{d\Upsilon}\right|_{\Upsilon=0}, \quad   \mathcal{P}_{cl}:=\lim_{\Upsilon\rightarrow0}  \mathcal{P}_{cl}.
\end{equation}
We repeat the above procedure, successively resolving all the commutators in $\mathcal{P}_{cl}$. Note that in the first application of the commutation rules, only the rules which do not involve $\hat E^P_\pm(v,\mu)$ can enter the game. 

After $\mathcal{P}_{cl}$ is free from commutators we need to replace all the commutators inside $\mathcal{P}_{qu}$. We shall do this successively by using the commutation rules with $\Upsilon=0$. We also  neither update $\mathcal{P}_{cl}$ nor $\mathcal{P}_{qu}$ at that stage. As a result of the whole first step of the algorithm, both objects  $\mathcal{P}_{cl}$ and $\mathcal{P}_{qu}$ are sums of monomials.

\subsubsection{Step II: Shifting the holonomies}
In the first part of this step we work with monomials inside $\mathcal{P}_{cl}$. We successively replace
\begin{equation}
\hat X \hat h_{ab}^{(j)} (v,\mu) \rightarrow \hat h_{ab}^{(j)} (v,\mu) \hat X +\Upsilon  [\hat X,\hat h_{ab}^{(j)} (v,\mu) ],
\end{equation}
where $\hat X$ can either be $\hat E_\pm$ or $\hat Q$. After each replacement we apply the update rule Eq. (\ref{update}) and continue the procedure until all the holonomies in $\mathcal{P}_{cl}$ are shifted to the left.

When the above procedure is over, the object $\mathcal{P}_{qu}$ is complete. We only need to resolve (without the update step) the new commutators which entered  $\mathcal{P}_{qu}$. B observation \ref{obs2} the order of the elements in $\mathcal{P}_{qu}$ is irrelevant.

\subsubsection{Step III: dealing with $\hat E_\pm$}
First, we work with the quantum corrections as this is much easier. We simply replace
\begin{align} \label{repQ}
\hat E^K_-(v,\mu)& \rightarrow \mathcal E\frac{2i\eta}{t}\delta^K_0,\\
\hat E^K_+(v,\mu)&  \rightarrow 0,
\end{align}
for all relevant instances inside $\mathcal{P}_{qu}$. This can be done due to being the leading order corrections according to observation \ref{obs1}.

Note that due to the above rule we can freely commute $\hat E_\pm$ with $\hat Q$ also inside $\mathcal P_{cl}$. This simplification is possible because even though a generic replacement $\hat E^P_\pm(v,\mu)\hat Q_{v'}^N \rightarrow \hat Q_{v'}^N \hat E^P_\pm(v,\mu)$ performed inside  $\mathcal P_{cl}$ would potentially bring to $\mathcal P_{qu}$ a contribution involving $[\hat E^P_\pm(v,\mu),\hat Q_{v'}^N]$, such terms do always vanish due to (\ref{repQ}). For $\hat E_-$ this is immediate as the commutator in question only involves $\hat E_+$. For $\hat E_-$ we  get the factor $\epsilon^{(n:\tilde\mu_1\tilde\mu_2\tilde\mu_0)}_{00K}f_{PK0}$, which after summing over $K$ leads to $-f_{P00}\equiv0$ due to (\ref{EpsilonTrick}).

We therefore use the above fact to freely shift all $\hat Q$ appearing inside $\mathcal P_{cl}$ to the right. Therefore, $\mathcal P_{cl}$ is a collection of monomials of the form "$\sim \hat h..\hat h\hat E..\hat E \hat Q..\hat Q$", where with $\sim$ we cover all non-operator factors to be contracted. For the quantum corrections we already obtained a simplified form "$\sim \hat h..\hat h \hat Q..\hat Q$".

For the monomials constituting $\mathcal P_{cl}$ we need to employ a more elaborate replacement:
\begin{align} \label{repC}
\hat E^K_\pm(v,\mu)& \rightarrow \mathcal E\left( R^K(v,\mu)\mp L^K(v-e_{\mu},\mu)  \right).
\end{align}
Afterwards we "abelian" shift the rightmost $L$ operator to the right (through all $R$), and in this particular position replace it according to the rule (\ref{replaceLtoR}):
\begin{equation}
L^{K}(v,m) \rightarrow e^{-i K z} R^{K}(v,m).
\end{equation}
We successively repeat this procedure until no $L$-type operators are left.

\subsubsection{Step IV: Link-splitting}
Before we will be able to perform the splitting, we first need to replace the holonomies by $D$ and $D^\dagger$ operators according to the formula (\ref{holonomies}). This is very important as the holonomy with different sign of $\mu$ gives $D$ or $D^\dagger$ on a different vertex.

After the above step is done we are also allowed to impose the periodic boundary conditions, replacing all vertices by its coordinates modulo $M$ (this step is essential when the computer-algebra-methods are involved). If need be, one shall also eliminate trivial products of holonomies according to the formula:
\begin{equation}
D^{(j)}_{sa}(v,\mu)\left[D^{(j)}_{ab} (v,\mu)\right]^\dagger\rightarrow \delta_{sb}/(2j+1).
\end{equation}
Finally, we use the identity $\left[D^{(j)}_{ab}\right]^{\dagger} = (-1)^{b-a} D^{(j)}_{-b-a}$ to replace all $D^\dagger$. 

At that stage both  $\mathcal P_{cl}$ and $\mathcal P_{qu}$ are collections of monomials, i.e.
\begin{equation}
\mathcal P_{cl/qu}=\sum_{s} \mathcal{P}_{cl/qu}^{(s)},
\end{equation}
with all  $\mathcal{P}_{cl/qu}^{(s)}$  being of  the form $\sim D..DR..R\hat Q..\hat Q$. We are therefore ready for the proper part of the link splitting. To this end we write
\begin{equation}
\mathcal{P}_{cl/qu} = \sum_s\prod_{v\in\mathbb{Z}^3_M}\left(\prod_{m=1}^3 \l_{cl/qu}^{(s)}(v,m)\right)\;\hat  Q_{v}^{N_{v}},
\end{equation}
where each $\l_{cl/qu}^{(s)}(v,m)$ is a monomial $\sim D..DR..R$ with all $D$  and $R$ taken at the same $(v,m)$. By $N_v$ we denote the collected power of $\hat Q_v$ at  $v$.

\subsubsection{Step V: Simplification of holonomies}
After the link splitting, for a fixed monomial and at a given $(v,m)$ we are left with the product of holonomies preceding the product of the $R$ operators. However, we can reduce every product of holonomies to a single $D$ operator. To this end we just recursively apply the formula
\begin{equation}
\label{DoubleHolProd}
D^{(j_1)}_{ab}(v,m)D^{(j_2)}_{cd}(v,m)=: \sum_{j_{tot}=|j_1-j_2|}^{j_1+j_2} Z_{j_1 j_2 j_{tot}}[a,b,c,d]  D^{(j_{tot})}_{-a-b-c-d}(v,m),
\end{equation}
where
\begin{equation}
Z_{j_1 j_2 j_{tot}}[a,b,c,d]=(2j_{tot}+1) (-1)^{a+c-b-d} \left(\begin{array}{ccc}
j_1 & j_2 & j_{tot}\\
a &  c & a+c
\end{array}\right)\left(\begin{array}{ccc}
j_1 & j_2 & j_{tot}\\
b &  d & b+d
\end{array}\right),
\end{equation}
which simply represents known rules for addition of spin in the fashion of the $SU(2)$ group \cite{BS68,Varshalovich}.

As has already been mentioned, knowing the maximal value of $j_{tot}$ appearing in the whole computation would make the task easy, though this is not a requirement of the algorithm.

\subsubsection{Step VI: Final replacements}
According to all previously described simplifications and after looking at the form we arrived at, we can conclude that the average value of the both the classical and the quantum contribution can be calculated with the help of the results from Observation \ref{obs5}. To this end we define
\begin{equation}
\hat F_{cl/qu}^{(s)}=\prod_{v\in\mathbb{Z}^3_M}\prod_{m=1}^3 \l_{cl/qu}^{(s)}(v,m),
\end{equation}
so that
\begin{equation}
\mathcal{P}_{cl/qu}^{(s)} =  \hat F_{cl/qu}^{(s)}\prod_{v\in\mathbb{Z}^3_M}\;\hat  Q_{v}^{N_{v}}.
\end{equation}
Therefore $\langle \Psi^t_\gamma,\hat F_{cl}^{(s)}\prod_v \hat Q_v^{N_v}\Psi^t_\gamma \rangle$ can directly be calculated from Eq. (\ref{Q_trick_quantum}), while for $\langle \Psi^t_\gamma,\hat F_{qu}^{(s)} \prod_v \hat Q_v^{N_v}\Psi^t_\gamma \rangle$ only the contribution from (\ref{Q_trick_classical}) is sufficient. At the end we need to sum all contributions (i.e. sum over $s$) together with their prefactors (not mentioned here explicitly). As stated at the beginning, all indices also need to be contracted.

%\begin{equation}
%\langle \Psi^t_\gamma,\hat H\Psi^t_\gamma \rangle = \sum_s \langle \Psi^t_\gamma,\hat F_{cl}^{(s)}\prod_v \hat Q_v^{N_v}\Psi^t_\gamma \rangle+t \sum_s \langle \Psi^t_\gamma,\hat F_{qu}^{(s)}\prod_v \hat Q_v^{N_v}\Psi^t_\gamma \rangle
%\end{equation}

\section{Expectation value of the euclidean scalar constraint}
\label{s4}

\subsection{Discretisation and Quantisation of the scalar constraint}
The scalar and diffeomorphism constraint encapsulate the dynamical content of general relativity in the continuum. However, in the present work we consider its discretisation on a spatial lattice and their action can not straightforwardly be lifted to the lattice, but needs to be approximated by the quantities of the lattice phase space.\footnote{Notice, that this is different to the situation of the Gauss constraint: (\ref{GaussConst}) and its flow on the phase space given by the vector field $\exp\left(\{\vec{G} [\vec{\Lambda}] ,\,\cdot\,\}\right)$ can be translated into a discrete version, given by $G^\epsilon_J(v)=\sum_{e;e\cap v\neq 0} s_{e_v} P^J(e_v)$, see \cite{wojtek} for further details.}
Many such discretisations are possible, however we will focus in this work on a single strategy and elucidate on some necessary choices. Moreover, this section considers only the euclidean part of the scalar constraint $C_E$ from (\ref{scalar_ce}) as the Lorentzian part works analogously.\\

First, we will employ Thiemann's identity \cite{Thi96_1,Thi96_2} to express:
\begin{align}
C_E = \frac{4}{\kappa^2 \beta}F^J_{ab}(A)\epsilon^{abc} \{V[\sigma],A^J_c\}
\end{align}
with $V[\sigma]=\int dx^3 \sqrt{|\det(E)|}$.  We can express this object now in terms of the  holonomies and fluxes of the lattice from (\ref{holonomy}) and (\ref{gaugecov_fluxes}) for edges of coordinate length $\epsilon$ in order to obtain an expression that reduces to $C_E(x=v)$ in the limit $\epsilon\to 0$: \cite{wojtek}
\begin{align}\label{disc_scalar_ce}
C_E^{\epsilon}(v):= \frac{-4}{\kappa^2\beta}\sum_{ijk\in L} \frac{\epsilon(i,j,k)}{T_v}tr\left((h(\square^{\epsilon}_{v,ij})-h^\dagger(\square^\epsilon_{v,ij})) h(e_k)\{h^\dagger(e_k), V^\epsilon[\sigma]\}\right)
\end{align}
with $L=\{1,2,3,-1,-2,-3\}$ labelling the different directions of the lattice and and $e_k$ indicating an edge starting from $v$ in direction $k$. Also $h(\square^\epsilon_{v,ij})$ indicated the minimal plaquette of the lattice starting at $v$ along direction $i$ and returning along direction $j$. Finally, for the cubic lattice $T_v=2^3=8$ and
\begin{align}\label{disc_volume}
V^\epsilon[\sigma]:= \sum_{v}\sqrt{|Q_v|/(6T_v)},\hspace{10pt}
Q_v= \sum_{e_i\cap e_j \cap e_k=v}\epsilon_{IJK} \epsilon(i,j,k) P^I(e_i)P^J(e_j)P^K(e_k)
\end{align}

In order to quantise this expression, we express holonomies and fluxes by their quantum operators (\ref{quantisation}), replace $i\hbar \{\cdot,\cdot\} \rightarrow [\cdot,\cdot]$ and use the Ashtekar-Lewandowski volume operator for (\ref{disc_volume}). Finally, since we are only interested in {\it expectation values including the next-to-leading order} we switch the latter one to the Giesel-Thiemann volume, ending up with the quantum operator:
\begin{align}\label{scalar_constraint_quant}
\hat{C}^\epsilon_E(v):=p_{\rm Euc} \sum_{ijk\in L}\left((\hat h(\square^{\epsilon}_{v,ij})-\hat h^\dagger(\square^\epsilon_{v,ij})) \hat h(e_k)\,[\hat h^\dagger(e_k), \hat{V}^{GT}_{2,v}+\hat{V}^{GT}_{2,v+e_k} ]\right)
\end{align}
with $p_{\rm Euc}=1/(i\kappa \beta l^2 t)$. A few remarks are due:
\begin{itemize}
\item The appearance of the additional $\hat{V}_{v+e_k}$ inside the commutator is different from other choices in the literature \cite{AQG1}. However, we take this choice in order to restore the discrete version (\ref{disc_scalar_ce}) in the classical limit, which we understand as the leading order of the expectation value in coherent states. Would we not have this term, we would get a discretised constraint on the lattice with only $V^\epsilon(v)$ in the Poisson bracket, {\it not} $V^\epsilon[\sigma]$. As it was shown in \cite{wojtek} if not at least one of the arguments of the Poisson bracket is invariant under the symmetries of the lattice, the symmetry restriction to the reduced phase space (spanned by coordinates $(p,c)$) does not work. I.e. to compute the dynamics of the scalar constraint in cosmology, one {\it cannot} first restrict and then compute its flow. The latter one is only possible if one argument of the Poisson bracket is indeed symmetric. However, this is a prerequisite of the {\it effective dynamics conjecture} which we will discuss in section \ref{s_effectivedynm}. Thus, in order to have a chance for it to work, we add the additional term $\hat{V}_{v+e_k}$.
\item We use the 2nd Giesel-Thiemann operator $\hat V^{GT}_{2,v}$ albeit being only interested in the first order $\mathcal{O}(t)$. This is due to the fact, that the volume appears in an commutator: here the $t^0$-order of the expectation value is indeed cancelling. The $t^1$-order becomes the classical order (yielding the classical term due to the $1/t$ in $p_{Euc}$) and the interesting next-to-leading order is indeed at $t^2$.
\end{itemize}

\subsection{Analytic computation of the expectation value of the euclidean scalar constraint}

In order to showcase the above described step-by-step guide, we have implemented it in a mathematica file. In this section, we will as an example highlight which steps the algorithm takes when evaluating the expectation value of cosmological coherent states for the euclidian part of the scalar constraint. We state the claim produced by the machine and afterwards verify it via a lengthy analytical computation, showcasing the strength of the compact algorithm and providing a consistency check.\\

The goal is to compute $\hat{C}_E^\epsilon$ from (\ref{scalar_constraint_quant}) in cosmological coherent states. Upon this tasks, our algorithm is doing the following:

{\bf I. Resolve commutators:} There are two commutator at classical order to resolve. After replacing $\hat{V}^{GT}\to c_N \hat{Q}^N$ we attack them using equation (\ref{Replacement_hQ}), e.g.
\begin{align*}
  &\hat{h}(v,3)[\hat{h}^\dagger(v,3),\hat{Q}_v^N] \rightarrow\\
  &-6i \mathcal{E} N \left(
  \epsilon^{(n:123)}_{ILK} [n\uptau n]^{(3,1/2,K)} \hat{E}^I_-(v,1) \hat{E}^L_-(v,2) \hat{Q}^{N-1}_v-
  \Upsilon\frac{N-1}{2}\hat{h}(v,3)[\hat A^{(j)}_{a b}(v,v,3),\hat Q_{v}]\hat Q^{N-2}_{v}\right)\\
  &\rightarrow
  -6i \mathcal{E} N \bigg(
  \epsilon^{(n:123)}_{ILK} [n\uptau n]^{(3,1/2,k)} \hat{E}^I_-(v,1) \hat{E}^L_-(v,2) \hat{Q}^{N-1}_v-\Upsilon\frac{N-1}{2}\epsilon^{(n:123)}_{I'L'K'}[n\uptau n]^{(3,1/2,K')}
  6i\mathcal{E}(N-1)\\
  &\hspace{30pt}[-\hat{A}^{(j)}(v,v,3)E^{I'}_-(v,1)E^{L'}_-(v,2) +\hat{B}^{I'}_{-}(v,1)\hat E^{L'}_-(v,2) +\hat{B}^{L'}_{-}(v,2) E^{I'}_-(v,1)
  ]\hat Q^{N-2}_{v}\bigg).
\end{align*}
{\bf II. Shift holonomies to the left:} We are in the fortunate situation that all holonomy operators are already on the left from the onset and therefore there is nothing to do.\\
{\bf III. Dealing with vector fields:} As discussed above, $\hat{E}_-$ and $\hat Q$ commute freely thus we can shift all of the $\hat{E}_-$ to the right and replace according to (\ref{repQ}) and (\ref{repC}), respectively. The latter one creates  left-invariant vector fields which we immediately turn into a right-invariant vector fields via (\ref{replaceLtoR}).\\
{\bf IV. Link splitting:}
One has to proceed with each of the terms separately. There are currently 48 in $\mathcal{P}_{cl}$ and 36 in $\mathcal{P}_{qu}$ of which we show one explicitly as example:
\begin{align}
     &\epsilon^{(n:123)}_{ILK}\langle\Psi^t_\gamma, (\hat{h}(\Box^\epsilon_{v,12})-\hat{h}^\dagger(\Box^\epsilon_{v,12})) [n\uptau n]^{(3,1/2,K)} R^I(v,1) (e^{izL} R^L(v,-1))\,\Psi^t_\gamma\rangle
     \\
     &\Rightarrow\nonumber\\
     &l_{(v,1)}= D^{(1/2)}_{ab} R^I,\;l_{(v+e_1,2)}=D^{(1/2)}_{bc},\;l_{(v+e_2,1)}=[D^{(1/2)}_{cd}]^\dagger ,\; l_{(v,2)}=[D^{(1/2)}_{de}]^\dagger, l_{(v-e_2,2)}= R^L.
\end{align}
{\bf V. Simplify products of holonomies:} There are no products of holonomies on a single edge appearing, therefore this step finishes after removing the complex conjugation on some Wigner matrices.\\
{\bf VI. Final Replacement:}
We replace each $l_e$ according to {(\ref{Q_trick_classical}) and (\ref{Q_trick_quantum}). As these are several terms we omit them due to lack of space and employ useage of a Mathematica implementation of the algorithm. It produces the following result:\\

{\bf Claim:} The expectation value of $\hat{C}^\epsilon_E(v)$ in cosmological coherent states $\Psi^t_\gamma$ from (\ref{cosmological_coherent_state}) parametrized by $\eta=2\epsilon^2p/(l^2\beta)$ and $\xi=\epsilon c$ is:
\begin{align}
&\langle\Psi^t_\gamma, \hat{C}_E^\epsilon(v)\,\Psi^t_\gamma\rangle=\frac{6}{\kappa}\epsilon\sqrt{p} \sin(\xi)^2\left[1+t\left(
-\frac{1}{4}-\frac{13{\rm coth}(\eta)}{8\eta}+\frac{11}{8\eta {\rm sinh}(\eta)}+\frac{9}{32\eta}
\right)
+t\frac{3i}{8}\frac{\sin(\xi/2)^2}{\sin(\xi)}\right]+\mathcal{O}(t^2).
\end{align}
\\

{\it Proof.} We will now supplement an analytic proof of the claim, thereby presenting a consistency check for the algorithm.
First, we make use of the fact that the expression on the lattice is invariant under rotations of ninety degrees and reflections, as well is the state $\Psi^t_\gamma$ which mimics an isotropic universe. Instead of all triples $i,j,k$ we will fix $i=+1,j=+2,k=+3$ and use that all  $3\times 2^3$ triples will result in the same expectation value and can then be summed by linearity.\\
We split our computation into two parts
\bal
\langle\Psi^t_\gamma, \hat{C}^\epsilon_E(v)\,\Psi^t_\gamma\rangle &= p_{\rm Euc}\frac{(\beta \hbar \kappa)^{3/2}}{2^{7/2}\sqrt{3t}}\;(3\times 2^3)\; ( \langle\Psi^t_\gamma, \hat{C}_E^{\epsilon,(1)}\,\Psi^t_\gamma\rangle + \langle\Psi^t_\gamma, \hat{C}_E^{\epsilon,(2)}\,\Psi^t_\gamma\rangle),\\
\langle \Psi^t_\gamma, \hat{C}_E^{\epsilon,(1)}\,\Psi^t_\gamma\rangle &= \sum_{N=0}^{10}c_N\, \langle\Psi^t_\gamma, (\hat{h}(\Box^\epsilon_{v,12})-\hat{h}^\dagger(\Box^\epsilon_{v,12}))\hat{h}(e_3)[\hat{h}^\dagger(e_3),\hat{Q}_v^N]\,\Psi^t_\gamma\rangle,\\
\langle\Psi^t_\gamma,  \hat{C}_E^{\epsilon,(2)}\,\Psi^t_\gamma\rangle &= \sum_{N=0}^{10}c_N\, \langle\Psi^t_\gamma, (\hat{h}(\Box^\epsilon_{v,12})-\hat{h}^\dagger(\Box^\epsilon_{v,12}))\hat{h}(e_3)[\hat{h}^\dagger(e_3),\hat{Q}_{v+e_3}^N]\,\Psi^t_\gamma\rangle.
\eal
First, we focus on $\hat{C}_E^{(1)}$: We denote by $\langle\hat F\rangle$ the expectation value of a polynomial $\hat{F}$ in the state $\psi^t_h$ with $h=\exp(-(\xi-i\eta)\underline{\tau}_3)$ that is $n=\tilde n= {\rm id}$. This short cut is used when considering all combinations of the product $\hat{Q}^N_v$ and extracting the $SU(2)$-elements $n_1,n_2,n_3$. This is done analogously to step IV. of the algorithm - Link splitting -, i.e.:
\bal
&\langle \Psi^t_\gamma, \hat{C}_E^{\epsilon,(1)}\,\Psi^t_\gamma\rangle =
\sum_{N=0}^{10}c_N\,(6i)^N \epsilon_{I_1,J_1,K_1}^{(n)}...\epsilon^{(n)}_{I_N,J_N,K_N} \sum_{s_1,s_2,s_3=0}^N \binom{N}{s_1}\binom{N}{s_2}\binom{N}{s_3}\times\\
&\times 
\bigg(
D^{(\frac{1}{2})}_{aa'}(n_1)
\langle \hat{h}_{a'b} R^{I_1}...R^{I_{s_1}} \rangle D^{(\frac{1}{2})}_{bb'}(n_1^\dagger n_2)
\langle \hat{h}_{b'c} \rangle 
D^{(\frac{1}{2})}_{cc'}(n_2^\dagger n_1)
\langle \hat{h}_{c'd}^\dagger \rangle D^{(\frac{1}{2})}_{dd'}(n_1^\dagger n_2 )
\langle \hat{h}_{d'e}^\dagger R^{J_1}...R^{J_{s_2}}\rangle D^{(\frac{1}{2})}_{ee'}(n_2^\dagger)
\nonumber\\
&\times \langle \hat{h}_{e'f} \left[ \hat{h}_{fa}^\dagger, R^{K_1}... R^{K_{s_3}} \right] \rangle-(n_1\leftrightarrow n_2)\bigg)
\langle R^{I_{s_1+1}}...R^{I_N} \rangle
\langle R^{J_{s_2+1}}...R^{J_N} \rangle
\langle R^{K_{s_3+1}}...R^{K_N} \rangle\nonumber.
\eal
Following this brute force computation, we observe that the  fourth line of (\ref{hRRR}) does not contribute in this case: were on a single edge a $\mathcal{O}(t)$-correction due to two right-invariant vector fields with non-zero indices, then (\ref{EpsilonTrick}) implies that also two other indices need to be non-vanishing, thus of non-leading order. But then the first edge and at least a second one combine to $\mathcal{O}(t^2)$ in which we are not interested - thus can be neglected. Similarly, after the commutator has been resolved, none of the remaining vector field with label $K$ can have a non-zero index, else (\ref{EpsilonTrick}) causes it to be of order $\mathcal{O}(t^2)$ due to requiring a non-zero index some other edge. The only remaining cases above are therefore (i) all indices equal to zero %note that the third line also has a contribution w/ s_A= 0.
or (ii) one index $I_A$ or $J_A$ is non-zero, where $R^{K_A}$ got absorbed by the commutator. Note that in the latter case - as well when considering the contribution from the third line with zero-index - we have to carefully sum over the allowed values. Explicitly, we obtain upon plugging into (\ref{hRRR}):
\bal
&\langle \Psi^t_\gamma, \hat{C}_E^{\epsilon,(1)}\,\Psi^t_\gamma\rangle =\sum_{N=1}^{10} (6i)^N c_N
(-)^N \sum_{s_1,s_2,s_3=0}^N \binom{N}{s_1}\binom{N}{s_2}\binom{N}{s_3}\\
&\Bigg(2\text{Im}\left[D^{(\frac{1}{2})}_{ae}(U_1U_2U_1^\dagger U_2^\dagger)\right](\gamma^{1/2}_{1/2})^4 \left(\frac{\eta i}{t}\right)^{3N-1}
\prod_{i=1}^2\left[1+\frac{t}{2\eta}\left(\frac{s_i(s_i+1)}{2\eta}-s_i \coth(\eta)\right)\right]\times\nonumber\\
&\;\;\;\;\times
\prod_{i=1}^3\left[
1+\frac{t}{2\eta}\left(\frac{(N-s_i)(N-s_i+1)}{2\eta}-(N-s_i) \coth(\eta)\right)\right]\nonumber
\times\\
&\;\;\;\;\times\left[
\sum_{A=1}^{s_3}[\uptau_{K_A}]_{ea}\delta_{k_A,0} \left[1+\frac{t}{2\eta}(\frac{(s_3-1)s_3}{2\eta}-(s_3-1)\coth(\eta))\right]+\frac{t\,i}{2\eta}\sum_{A<B}^{s_3}[\uptau_{K_A}\uptau_{K_B}]_{ea}\delta_{K_A,0} \delta_{K_B,0}\right]\nonumber
\\
%This is the term with all zero contributions from the third line. It does not matter which of the s_1/s_2 many Rs contributes to it.
&+(-\epsilon^{(n)}_{000})\left(\frac{\eta i}{t}\right)^{3N-1}\, s_3\,\sum_{A=0}^{s_1}\,[ \underline{\tau}_{3}]_{ea} \frac{t\,i}{2\eta}\left[ D^{(\frac{1}{2})}_{ae}(\underline{\tau}_1 U_{12}-U_{21} n_1 \overline{\uptau}_{0} n_1^\dagger) +
D^{(\frac{1}{2})}_{ae}(U_{12} n_2 \overline{\uptau_{0}}n_2^\dagger- \underline{\tau}_2 U_{21} ) \right]\Bigg)\nonumber\\
%We choosen s_3 many of N to go in +3. Then s_3 combination for the triple to pick, which fixes +1 in order for this term to exist. Then remain N-1 many R to freely pick \pm1 directions
&+\sum_{N=1}^{10}(6i)^Nc_N(-)^N 2^N\sum_{s_3}^N \sum_{s=0}^{N-1}\binom{N}{s_3}s_3\binom{N-1}{s}
\left(\frac{\eta i}{t}\right)^{3N-1}\frac{t\,i}{2\eta}[\underline{\tau}_{K_A}]_{ea}\nonumber\\
&\;\;\;\;\times \left[(-\epsilon^{(n)}_{I_A,0,K_A})(1-I_A\,{\rm tanh}(\frac{\eta}{2})) D^{(\frac{1}{2})}_{ae}(n_1 [\uptau_{I_A}] n_1^\dagger U_{12}-U_{21} n_1 [\overline{\uptau}_{I_A}] n_1^\dagger)\delta^{I_A}_{\pm 1}
+\right.\nonumber\\
&\hspace{80pt}+\left.(-\epsilon^{(n)}_{0,J_A,K_A})
(1-J_A\,{\rm tanh}(\frac{\eta}{2}))
D^{(\frac{1}{2})}_{ae}(U_{12} n_2 [\overline{\uptau_{J_A}}]n_2^\dagger-n_2 [\uptau_{J_A}] n_2^\dagger U_{21} )\delta^{J_A}_{\pm 1} \right]+\mathcal{O}(t^2),
\nonumber
\eal
where $U_{ij}:=U_iU_jU_i^\dagger U^\dagger_j$ with $U_I$ from (\ref{classical_hol}), also recall $\gamma_{1/2}^{1/2}=-\gamma^{1/2}_{1/2}$ and:
\bal
2\text{Im}(A):=A-A^\dagger,\hspace{20pt}
[\hat{h}_{ab},R^K]=[\uptau_K]^{(1/2)}_{ac}\hat{h}_{cb},\hspace{20pt} 
D^{(1)}_{-K-L}(g)[\uptau^L]_{mn}^{(j)} = D^{(j)}_{mm'}(g^\dagger)[\uptau^K]_{m'n'}^{(j)}D^{(j)}_{n'n}(g).\nonumber
\eal
We note that $tr(U_{ij}-U_{ji})=0$ to get:
\bal
&\langle \Psi^t_\gamma, \hat{C}_E^{\epsilon,(1)}\,\Psi^t_\gamma\rangle =
\sum_{N=1}^{10} (6i)^N c_N (-)^N \,tr(U_{12}\underline{\tau}_3-U_{21}\underline{\tau}_3)\left(\frac{\eta i}{t}\right)^{3N-1}\,2^{2N} \sum_{s_3=0}^N \binom{N}{s_3} s_3\nonumber
\\
&\bigg[1+ \frac{t}{2\eta}(-\frac{\eta}{2}-2{\rm tanh}(\frac{\eta}{2})-(3N-1){\rm coth}(\eta))\nonumber
\\
&+\frac{t\,2^{-N}}{4\eta^2}\sum_{s=0}^N\binom{N}{s}(2(N-s)(N-s+1)+2s(s+1)+(N-s_3)(N-s_3+1)+s_3(s_3-1))\nonumber
\\
&+\frac{t\,i}{\eta}2^{-N-1} \sum_s \binom{N}{s}\,s\,tr(\underline{\tau}_3\underline{\tau}_1 U_{12}+\underline{\tau}_3U_{21}\underline{\tau}_1-\underline{\tau}_3U_{12}\underline{\tau}_2-\underline{\tau}_3\underline{\tau}_2 U_{21})\bigg]\nonumber
\\
&-\sum_{N=1}^{10}(6i)^Nc_N(-)^N 2^{N}\sum_{s_3}^N \sum_{s=0}^{N-1}\binom{N}{s_3}s_3\binom{N-1}{s}\left(\frac{\eta i}{t}\right)^{3N-1}\frac{t\,i}{\eta 2}(1-j\, {\rm tanh}(\frac{\eta}{2}))\;\delta^J_{\pm 1}\times\nonumber\\
&\;\;\;\;\times tr\left(\epsilon^{(n)}_{J,0,K}\underline{\tau}_K (n_1 \uptau_{J} n_1^\dagger U_{12}+U_{21} n_1 \uptau_{J} n_1^\dagger)
-\epsilon^{(n)}_{0,J,K}\underline{\tau}_K 
(U_{12} n_2 \uptau_{J}n_2^\dagger+n_2 \uptau_{J} n_2^\dagger U_{21})
\right)+\mathcal{O}(t^2).
\eal
A calculation of the traces, e.g. in Mathematica, returns
%\begin{align*}
%tr^{(1/2)}[U_{12}\underline{\tau}_3-U_{21}\underline{\tau}_3]=-\sin(\xi)^2
%\end{align*}
%and:
%tr( \tau_3\tau_1 U_{12}+\tau_3U_{21}\tau_1-\tau_3U_{12}\tau_2-\tau_3\tau_2 U_{21}) =  2 \sin(\xi/2)^2 \sin(\xi)
%and the other:
%\bal
%tr(\epsilon...)=-i \sin(\xi)^2 tanh(\eta/2)-2 \sin(\xi/2)^2\sin(\xi)
%\eal
%Then:
%
%  \sum_{s=0}^{n} \binom{n-1}{s} s  = 2^{n-1} n
%
%  \sum_{s=0}^{n-1} \binom{n-1}{s}   = 2^{n-1}
%
% \sum_{s=0}^N \binom{n}{s} ((n - s) (n - s + 1) + s (s + 1)) = 2^{n-1} n(n+3)
%
%  \sum_{s=0}^N \binom{n}{s}  s   ((n - s) (n - s + 1) + s (s + 1)) = 2^{n-2}n(n+2)(n-1)
%

\bal
&\langle \Psi^t_\gamma, \hat{C}_E^{\epsilon,(1)}\,\Psi^t_\gamma\rangle =
\sum_{N=1}^{10} (6i)^N c_N
(-)^N \left(\frac{\eta i}{t}\right)^{3N-1}\,2^{3N-1} \Bigg(N [-\sin(\xi)^2]\times
\\
&\;\;\;\times\left[1+\frac{t}{2\eta}\left(-\frac{\eta}{2}-2\tanh\left(\frac{\eta}{2}\right)-(3N-1)\coth(\eta)+\frac{1}{2\eta}N(N+3)+\frac{1}{4\eta}(N+2)(N-1)\right)
\right]+\nonumber
\\
&+\frac{t\,i}{\eta}\frac{N^2}{4} [2 \sin(\xi/2)^2 \sin(\xi)]-\frac{t\,i}{\eta}\frac{N}{4}[-i \sin(\xi)^2 {\rm tanh}(\eta/2)-2 \sin(\xi/2)^2\sin(\xi)] )\Bigg)\nonumber
\\
&=-i\sqrt{3\eta}\sin(\xi)^2\left[1-\frac{t}{4}-t(1-\frac{1}{4})\frac{{\rm tanh}(\eta/2)}{\eta}-\frac{t}{4\eta}{\rm coth}(\eta)+\frac{t}{8\eta^2}\frac{9}{4}\right]-i\sqrt{3\eta}\frac{t\,i}{2\eta}\sin(\xi/2)^2\sin(\xi)(\frac{1}{2}%note that the N^2-sum has additional 1/2 
+1)
\nonumber\\
&=-i\sqrt{3\eta}\sin(\xi)^2\left[1+t(-\frac{1}{4}-\frac{{\rm coth}(\eta)}{\eta}+\frac{3}{4\eta {\rm sinh}(\eta)}+\frac{9}{32\eta^2})\right]-i\sqrt{3\eta}\,t\, (\frac{3i}{4} \sin(\xi/2)^2\sin(\xi))+\mathcal{O}(t^2).\nonumber
\eal
%Interestingly: it does not make a difference whether we use \sum_{N=1}^6 or \sum_{N=1}^10 (w/ the respective coefficients)

There is still $\hat C^{(2)}_E$ to compute: It deviates from the previous starting point in the shifted vertex, at which the volume acts, $V^{GT}_{2,v+e_3}$. That is, there can be {\it no} quantum contributions due to mixing of right-invariant vector fields and the holonomy loop $\hat h (\square^\epsilon_{v,12})$. All other contributions can repeat. However, on the edge in direction +3 starting at $v$, we have the commutator of $\hat{h}$ and {\it left-invariant} vector fields.\\
We will bring this contribution into a form maximally close to the one known so far (as always neglecting contributions of higher order than $t$):
\bal
&\frac{1}{s_3}\langle \hat{h}_{ab}\left[\hat{h}_{bc},L^{k_1}...L^{k_{s_3}}\right]\rangle=\langle \hat{h}_{ab}\tau^0_{bb'}\hat{h}^\dagger_{b'c} L^0...L^0\rangle =\nonumber
\\
&\langle \hat{h}^{(0)}_{00}R^0...R^0\rangle\left(
\begin{array}{ccc}
1/2 & 1/2 & 0\\
a & -c & 0
\end{array}\right)\left(\begin{array}{ccc}
1/2 & 1/2 & 0\\
b & -b' & 0
\end{array}\right)\tau^0_{bb'}(-)^{c-b'}+\nonumber\\
&\hspace{40pt}+3\langle \hat{h}^{(1)}_{mn}R^0...R^0\rangle\left(
\begin{array}{ccc}
1/2 & 1/2 & 1\\
a & -c & -m
\end{array}\right)\left(
\begin{array}{ccc}
1/2 & 1/2 & 1\\
b & -b' & -n
\end{array}\right)\tau^0_{bb'}(-)^{-m+n+c-b'}=\nonumber
\\
&=\delta_{ac}\frac{1}{2^2}\sum_{b=-\frac{1}{2}}^{\frac{1}{2}}(-)^{1/2+a+1/2+b}(-i/2)(-)^{1/2-b}(-)^{c-b}\langle R^0..R^0\rangle+\nonumber\\
&\hspace{40pt}+3\left(\begin{array}{ccc}
1/2 & 1/2 & 1\\
a &-c & -m
\end{array}\right)\left(\begin{array}{ccc}
1/2 & 1/2 & 1\\
b & -b & 0
\end{array}\right)(-i/2)(-)^{1/2-b-m+c-b}\langle \hat{h}^{(1)}_{m0}R^0...R^0\rangle\nonumber
\\
&=0+3\left(\begin{array}{ccc}
1/2 & 1/2 & 1\\
a & -c & -m
\end{array}\right)\frac{b}{\sqrt{3/2}}(-)^{3/2-b}(-\frac{i}{2})(-)^{1/2-2b-m+c}\left(\frac{i\eta}{t}\right)^n\left(
\delta_{m0}\frac{t}{2\eta}\sum_{i=0}^{n-1}\eta^{-i}\partial_\eta \eta	^i\right) (1-t\frac{\tanh(\eta/2)}{\eta})\nonumber
\\
&=2(-ia/2)\delta_{ac}(-)^{3/2-3b+1/2-b}\left(\frac{i\eta}{t}\right)^n
(1-t\frac{\tanh(\eta/2)}{\eta})\left[1+\frac{t}{2\eta}(\frac{n(n+1)}{2\eta}-n\coth(\eta)\right]\nonumber
\\
&=\tau^0_{ac}\langle R^0... R^0\rangle (1-t\frac{\tanh(\eta/2)}{\eta}),\label{vpe3_trick}
\eal
where we used $SU(2)$-recoupling and that
\bal
\left(\begin{array}{ccc}
j & j & 0\\
m & n & 0
\end{array}\right)=\frac{(-)^{j+m}}{d_j}\delta_{m,-n},\hspace{30pt}
\left(\begin{array}{ccc}
1/2 & 1/2 & 1\\
m & -m & 0
\end{array}\right)=\frac{m}{\sqrt{3/2}}(-)^{3/2-m}.
\eal
Then, we use (\ref{vpe3_trick}), together with the same steps as before, in
\bal
&\langle \Psi^t_\gamma, \hat{C}_E^{\epsilon,(2)}\,\Psi^t_\gamma\rangle =
\sum_{N=0}^{10}c_N\,(6i)^N \epsilon_{I_1,J_1,K_1}^{(n)}...\epsilon^{(n)}_{I_N,J_N,K_N} \sum_{s_1,s_2,s_3=0}^N \binom{N}{s_1}\binom{N}{s_2}\binom{N}{s_3}\times\\
&
\bigg(
D^{(\frac{1}{2})}_{aa'}(n_1)
\langle \hat{h}_{a'b} \rangle D^{(\frac{1}{2})}_{bb'}(n_1^\dagger n_2)
\langle \hat{h}_{b'c} \rangle 
D^{(\frac{1}{2})}_{cc'}(n_2^\dagger n_1)
\langle \hat{h}_{c'd}^\dagger \rangle D^{(\frac{1}{2})}_{dd'}(n_1^\dagger n_2 )
\langle \hat{h}_{d'e}^\dagger \rangle D^{(\frac{1}{2})}_{ee'}(n_2^\dagger)\langle \hat{h}_{e'f} \left[ \hat{h}_{fa}^\dagger, L^{K_1}... L^{K_{s_3}} \right] \rangle_z
\nonumber\\
&\hspace{30pt}-(n_1\leftrightarrow n_2)\bigg) 
\langle R^{I_1}...R^{I_{s_1}} \rangle
\langle R^{I_{s_1+1}}...R^{I_N} \rangle
\langle R^{J_1}...R^{J_{s_2}} \rangle
\langle R^{J_{s_2+1}}...R^{J_N} \rangle
\langle R^{K_1}...R^{K_{s_3}} \rangle
\langle R^{K_{s_3+1}}...R^{K_N} \rangle\nonumber
\\
&=
\sum_{N=1}^{10} (6i)^N c_N
(-)^N \left(\frac{\eta i}{t}\right)^{3N-1}\,2^{3N-1}\;N [-\sin(\xi)^2]\times\nonumber
\\
&\;\;\;\times\left[1+\frac{t}{2\eta}\left(-\frac{\eta}{2}-2\tanh\left(\frac{\eta}{2}\right)-(3N-1)\coth(\eta)+\frac{1}{2\eta}N(N+3)+\frac{1}{4\eta}(N+2)(N-1)\right)
-t\frac{{\rm tanh}(\eta/2)}{\eta}\right]\nonumber\\
&=-i\sqrt{3\eta}\sin(\xi)^2\left[1-\frac{t}{4}-2t\frac{{\rm tanh}(\eta/2)}{\eta}-\frac{t}{4\eta}{\rm coth}(\eta)+\frac{t}{8\eta^2}\frac{9}{4}\right]+\mathcal{O}(t^2).
\nonumber
\eal
Finally, plugging everything together we end up with\footnote{Note that the appearance of the imaginary part is simply due to $\hat{C}^\epsilon_E$ not being self-adjoint. A common extension in the literature is therefore to work with $\hat{C}^\epsilon_E+(\hat{C}^\epsilon_E)^\dagger$, whose expectation value follows straightforward from this result.}
\begin{align}
&\langle\Psi^t_\gamma, \hat{C}^\epsilon_E(v)\,\Psi^t_\gamma\rangle = p_{\rm Euc}\frac{(\beta \hbar \kappa)^{3/2}}{2^{7/2}\sqrt{3t}}\;24 ( \langle\Psi^t_\gamma, \hat{C}_E^{\epsilon,(1)}\,\Psi^t_\gamma\rangle + \langle\Psi^t_\gamma, \hat{C}_E^{\epsilon,(2)}\,\Psi^t_\gamma\rangle)\\
&=p_{\rm Euc}\frac{(\beta \hbar \kappa)^{3/2}}{2^{7/2}\sqrt{3t}}\;24\;(-i\sqrt{3\eta}\sin(\xi)^2)\times\nonumber
\\
&\hspace{40pt}\times
\left[1+t\left(
-\frac{1}{2}-2\frac{{\rm tanh}(\eta/2)}{\eta}-\frac{{\rm coth}(\eta)}{4\eta}+\frac{9}{16\eta^2}-\frac{{\rm coth}(\eta)}{\eta}+\frac{3}{4\eta {\rm sinh}(\eta)}
\right)
+i\,t\frac{3}{4}\frac{\sin(\xi/2)^2}{\sin(\xi)}\right]+\mathcal{O}(t^2)
\nonumber\\
&=p_{\rm Euc}\frac{(\beta \hbar \kappa)^{3/2}}{2^{7/2}\sqrt{3t}}\;48(-i\sqrt{3\eta}\sin(\xi)^2)\left[1+t\left(
-\frac{1}{4}-\frac{13{\rm coth}(\eta)}{8\eta}+\frac{11}{8\eta {\rm sinh}(\eta)}+\frac{9}{32\eta^2}
\right)
+t\frac{3i}{8}\frac{\sin(\xi/2)^2}{\sin(\xi)}\right]+\mathcal{O}(t^2).\nonumber
\end{align}
\qed

\section{On the fate of the effective dynamics program of Loop cosmology}
\label{s_effectivedynm}

The effective dynamics program/conjecture is a tool often used in the LQC literature \cite{Boj05,Ash08,APS06a,Tav08,MRR18,KSW20,GMP20,AOS20}. It makes assumptions over the dynamics of the expectation values of certain observables (typically the volume of the universe). Without loss of generality we will talk in this section about the flow of a physical Hamiltonian, which for example can be realized in General Relativity  by deparametrizing with suitable dust reference fields \cite{dep7}.\\
The effective dynamics conjecture has several levels, which can be summarized as follows:
\begin{itemize}
\item[(A)] The state stays sharply peaked under dynamical evolution with respect to some Hamiltonian for the considered observable.
\item[(B)] The trajectory of the expectation value of the observable follows the trajectory of an effective Hamiltonian, which is the expectation value of the quantum Hamiltonian, evolved on the classical phase space.
\item[(C)] One can restrict the effective Hamiltonian and its flow on the classical phase space to a small subspace of  ``relevant'' degrees of freedom. On this subspace the restricted Hamiltonian and its flow with respect to the reduced Poisson brackets matches the evolution of the full effective Hamiltonian.
\end{itemize}
First, we want to stress that the program has been verified with great success in certain quantum models of the isotropic sector of cosmology, i.e. Loop Quantum Cosmology (LQC) \cite{APS06a,ADLP18}. Here, one deals with a quantum theory of a single degree of freedom, thus the last bullet point (C) of the above mentioned list becomes void. 
The classical phase space is parametrized by two parameters $p,c$ and endowed with the symplectic structure $\{p,c\}=\kappa\beta/6$. A quantisation of a symmetry restricted scalar constraint, coupled to a scalar field as clock content, could be analytically investigated: its expectation value in coherent spaces peaked sharply on classical values $p,c$ can be computed as some function $C_{cos}(c,p)$. Now, also its quantum flow on the corresponding coherent state was directly computed and verified to agree with the flow of the $C_{cos}(c,p)$ with the above mentioned Poisson bracket. Consequently, trust for the effective dynamics program increased and it was applied in other LQC-like scenarios as well, without being thoroughly proven.\\

Of course, there is a downside to the LQC quantisation strategy of only a single degree of freedom, i.e the scale factor: it is unclear how its relation to an actual field theoretical quantisation of GR does look like. In this section, we explore a bit in this direction, as - albeit not having access to a quantum field theory - we have a quantum theory with many degrees of freedom on the lattice.\\
Therefore, the third bullet point becomes paramount: in which situations is it possible to restrict the full dynamics of a Hamiltonian on the lattice phase space to some subspace, without loosing information? In \cite{wojtek}, this procedure was proven on a classical level to work, given the reduced phase space consists of those points that are left invariant by the action of some group of symplectomorphims and if the Hamiltonian is invariant under action of the same group as well. In \cite{wojtek} it was presented how gravity on the graph can be symmetry restricted to the phase space of isotropic cosmology parametrised by $p,c$ as outlined before. However, this  only proves the validity of (C) in the limit $t\to 0$. This section shall therefore consider, how it fares if the first order in $t$ is included and thus elucidates on the fate of the effective dynamics program.\footnote{We emphasis that in \cite{wojtek} invariance under the action of the group is necessary in order to reduce Poisson brackets. Thus, it is necessary in (\ref{scalar_constraint_quant}) to include the contribution of all touching vertices if we want to have a chance for the effective dynamics program to work.}\\

The validity of restricting the dynamics can be tested at the infinitesimal level with the toy model outlined in this paper so far: instead of investigating the whole flow of the Hamiltonian, we ask whether at small times the change of the expectation value of a time-evolved observable agrees with the classical flow induced on the classical observable on the restricted phase space.\\
As this is a necessary condition for the effective dynamics program, it suffices to show that this is in general not the case by considering the following: we take the euclidean part of the scalar constraint \footnote{This could be imagined as a toy model Hamiltonian if deparametrised with Gaussian dust \cite{dep7} and choosing the Barbero-Immirzi parameter $\beta	=i$.} as generator of the flow and the volume of the whole spatial manifold as observable. We choose an isotropic and homogeneous lapse function $\mathcal{N}(v)=1$, denoting $\hat{C}^\epsilon_E[1]=\sum_v \hat{C}^\epsilon_E(v) \mathcal{N}(v)$ and a lattice $\gamma$ of $N$ vertices along each direction. Also, we fix the total coordinate length of the torus to be 1 such that $\epsilon N=1$.
With the computer algorithm described in this paper it is now easy to compare the following two quantities:
\begin{align}
\langle \Psi^t_\gamma,\frac{1}{i\hbar} \big[ \hat{C}^\epsilon_E [1] \, ,\, \sum_v \hat{V}^{AL}\big] \,\Psi^t_\gamma \rangle
\hspace{30pt}\text{and}\hspace{30pt}
\big\{ \langle \Psi^t_\gamma,  \hat{C}^\epsilon_E[1] \,\Psi^t_\gamma \rangle\, , \, \langle \Psi^t_\gamma,  \sum_v \hat{V}^{AL}\,\Psi^t_\gamma \rangle \big\}.
\end{align}
Equality between both quantities is already known up to order $\mathcal{O}(t)$, but so far nothing is known when including the next-to-leading order.\\
The left side is (with the help of Mathematica) evaluated to be (recall that $\eta=\frac{2 p \epsilon ^2}{\beta  \ell^2}$)
\bal
&\langle \Psi^t_\gamma,\frac{1}{i\hbar} \big[ \hat{C}^\epsilon_E [1] \, ,\, \sum_v \hat{V}^{AL}\big] \,\Psi^t_\gamma \rangle=
-N^3 \frac{1}{2} 3 \beta  p \epsilon ^2 \sin (2 c \epsilon )+\\
&+t\,N^3\frac{3 \beta   \sin (c \epsilon )}{1024 p \epsilon ^2 \sinh(\eta)} \bigg[
4 \beta  l^2 p \epsilon ^2 \sinh \left(\frac{\eta }{2}\right) (-4 \cos (c \epsilon )+\cos (2 c \epsilon )-1)+16 \beta  \ell^2 p \epsilon ^2 (\cosh (\eta ) (57 \cos (c \epsilon )-1)-\nonumber\\
&\hspace{40pt}-41 \cos (c \epsilon )+1)-\sinh (\eta ) \left(\cos (c \epsilon ) \left(141 \beta ^2 \ell^4-256 p^2 \epsilon ^4\right)+\beta  \ell^2 \left(5 \beta  l^2+96 i p \epsilon ^2 \sin (c \epsilon )\right)\right)
\bigg]+\mathcal{O}(t^2).\nonumber
\eal
With the result from section \ref{s4} and the formula of the volume, e.g. from \cite{DL17b}, the right side is easily calculated. However, while both expressions cancel in their leading order, the same is {\it not true when including quantum corrections}:

\begin{align}
&\langle \Psi^t_\gamma,\frac{1}{i\hbar} \big[ \hat{C}^\epsilon_E [1] \, ,\, \sum_v \hat{V}^{AL}\big] \,\Psi^t_\gamma \rangle-
\big\{ \langle \Psi^t_\gamma,  \hat{C}^\epsilon_E[1] \,\Psi^t_\gamma \rangle\, , \, \langle \Psi^t_\gamma,  \sum_v \hat{V}^{AL}\,\Psi^t_\gamma \rangle \big\}=
\nonumber\\
&=\frac{3 \beta  t}{1024 p \epsilon ^5} \bigg[-48 i \beta  l^2 p \epsilon ^2 (\cos (\xi )-\cos (2 \xi ))+4 \text{csch}^2(\eta ) \sin (2 \xi ) \left(-\beta ^2 l^4+\cosh (2 \eta ) \left(\beta ^2 l^4-16 p^2 \epsilon ^4\right)\right)+
\nonumber\\
&+4 \text{csch}^2(\eta ) \sin (2 \xi ) \left(4 \beta  l^2 p \epsilon ^2 (22 \sinh (\eta )-15 \sinh (2 \eta ))+80 p^2 \epsilon ^4\right)-
\nonumber\\
&+\frac{ \sin (\xi )}{\sinh(\eta)} \left(4 \beta  l^2 p \epsilon ^2 \sinh \left(\frac{\eta }{2}\right) (-4 \cos (\xi )+\cos (2 \xi )-1)+16 \beta  l^2 p \epsilon ^2 (\cosh (\eta ) (57 \cos (\xi )-1)-41 \cos (\xi )+1)-\right.
\nonumber\\
&\left.-\sinh (\eta ) \left(\cos (\xi ) \left(141 \beta ^2 l^4-256 p^2 \epsilon ^4\right)+\beta  l^2 \left(5 \beta  l^2+96 i p \epsilon ^2 \sin (\xi )\right)\right)\right)\bigg]
+\mathcal{O}(t^2)
\nonumber\\
&\neq 0.
\end{align}

This shows, that upon including the quantum corrections, it becomes non-sensible to trust the effective dynamics program. The additional $\mathcal{O}(t)$ corrections if understood as a classical phase space functions do {\it not} provide reliable insight into the real quantum dynamics.\\
Therefore, as long as the quantum corrections of the spread are small compared against the discretisation effects due to the lattice, the effective dynamics conjecture has still a chance to find application (although item (A) and (B) remain to be carefully investigated even in the limit $t\to 0$). However, if one is interested in the effects that appear due to finite $t$ parameters, those must be investigated with other methods instead of the effective dynamics program!

\begin{remark}
One may propose to get at least a first intuition on the small time effects that a self-adjoint quantum Hamiltonian $\hat H$ enforces on some observable $\hat{O}$, by looking at the power series expansion of $e^{-is\hat{H}} \hat O e^{is\hat{H}}$. That is to investigate the expectation values of the $n$th-momenta $[\hat{H},\hat V]_{(n)}$ with the iterated commutator. However, such a procedure can only be trustful, if suitable convergence properties of the power series can be shown.
\end{remark}
\begin{remark}
Currently, the corrections which drive the dynamics in models such as LQC \cite{APS06a,ADLP18} are mainly driven by ambiguous discretisation choices. Although these feature a bounce, one can see by our investigations that said bounce is not due to quantum effects but due to discretisation artefacts. In other words, the bouncing behaviour can only be accepted if one takes the premise of a fundamentally discrete spacetime for granted. Conversely, one is typically interested in quantum field theories where the lattice regulator is removed, $\epsilon\to 0$. Next to the fact that this worsens the above situation for the effective dynamics program as soon as $t>> \epsilon$, the above framework does {\it not} give reliable results in this limit: In the context of Quantum Gravity the replacement of the Ashtekar-Lewandowski volume with the Giesel-Thiemann volume was necessary for the algorithm to work. However, said replacement is only valid in the limit $\epsilon>> t$, therefore one should not interpret the $\epsilon\to0$-limit of the above stated expressions as containing valid physical information.
\end{remark}

\section{Conclusions and Outlook}
\label{s_concl}
\numberwithin{equation}{section}
This paper presented a step-by-step `computer algorithm' for expectation values of polynomial operators in cosmological coherent states including the first order in the spread of the states. The cosmological coherent states are a tensor product of suitable gauge coherent states (GCS) which are suitable for all LGTs from \cite{TW1,TW2,TW3}. These GCS are labeled by classical phase space data $p,c$ of isotropic flat cosmology and sharply peaked in the sense, that the expectation value of any operator, corresponding to a classical function on the phase space, results in precisely the evaluation of its classical function in said phase space data in zeroth order of the spread of the states $t$. However, the first order is far from non-trivial, but captures important modifications due to the quantum nature of space time. In order to compute those, our algorithm enters the stage:\\
As input one can investigate any polynomial operator in the basic variables (holonomies and fluxes). Due to several crucial observations the computation can be drastically simplified, presenting a further step towards bringing LQG into the realm of computability. Complicated operators such as the volume operator can be tackled by manipulations such as the replacement from \cite{AQG2} and then observing that products of $\hat Q_v$ decouple from the remaining operator. The final simplifications were due to the observations from \cite{DL17b} that products of right-invariant vector fields contribute to the classical order when all indices are zero, and to the first order with at most one pair of non-zero indices. This removed the major part of contractions that need to be summed over, once the formula of \cite{LZ20} is employed, which translates the expectation value into a function of $p,c,t$.\\

A few words of caution have to be made at this point: the coherent states are of kinematical nature in the sense that - albeit their exceptional peakedness properties - they are not in the kernel of a known quantisation of the constraints, neither the scalar/diffeomorphism constraint nor the Gauss constraint. Next to considering different proposals (e.g. see \cite{TZ16,CFLS20}), it became popularized in Quantum Gravity approaches on the lattice, to utilize the so called group averaging procedure \cite{Thi09,Mar99,Mar00}, by which e.g. a simple tensor product over coherent states can be projected to a gauge invariant state \cite{Bahr}. It is worthwhile to note, that the expectation value of gauge-invariant operators, like the volume, do not get afflicted by this projection in its leading order in $t\sim \hbar$, but in its next-to-leading order. In future, one may adapt the present algorithm to include the additional corrections stemming from such a group averaging procedure. See the appendix of \cite{DL17b} for a strategy how these corrections can in general be computed.\\
Next, while it may be tempting to straightforwardly compute the expectation value of the scalar constraint (note that in this paper we have only shown the Euclidean part as a toy model, while the Lorentzian part can be computed analogously albeit with more computational effort) and to extract physical predictions from it, we emphasise again that the framework of LQG is plagued by many quantisation ambiguities stemming from several choices of how to approximate the continuum scalar constraint $C$ as a function on the lattice $C^\epsilon$ \cite{Thi96_1,Thi96_2,Warsaw15}. In the presence of a finite lattice, these ambiguities will overshadow the quantum corrections - as they appear on the leading order level - and therefore need to be dealt with first. Neither is any known quantisation of the scalar constraint cylindrical consistent which is the necessary condition to promote it to a continuum operator, nor can one simply take the limit of $\epsilon\to 0$ as the replacement of the Ashtekar-Lewandowski volume with the Giesel-Thiemann volume only works at finite lattice spacing $\epsilon>>t$. These issues will be attacked in future publications which now gain the support of being easier testable due to our algorithm.\\

Leaving the above mentioned technicalities aside, even at the current stage, one can use the algorithm to learn something about the dynamics of LQG: namely, we investigated in section \ref{s_effectivedynm} the fate of the {\it effective dynamics program}. In Loop Cosmology-like models of reduced systems, one often takes the expectation value of the scalar constraint in said system (or a function which is hoped to approximate it at leading order in $t$) and evolves physical quantities with respect to the flow generated by said expectation value due to the classical Poisson brackets. We asked whether such a strategy can have the potential to be viable once the many degrees of freedom of the lattice are considered and found the answer to be in the negative once the next-to-leading order is considered! A quick consistency check revealed that - although agreeing in the leading order - the $\mathcal{O}(t)$-modifications of $\langle [\cdot,\cdot]\rangle$ are different then those of $\{ \langle \cdot\rangle,\langle \cdot\rangle\}$ even for the simple case isotropic, flat cosmology on the lattice. This presents the infinitesimal level of the quantum evolution and is providing a correction that is of {\it the same order} as the new linear $t$ corrections whose influence one wants to investigate. Therefore, the effective dynamics conjecture is invalid for the quantum corrections of lattice LQG and instead one should rather be interested in a power series of higher order momenta of the quantum observables in question - which poses a problem that can be addressed with the present algorithm. We reserve this part for future research.

\vspace{1cm}
\noindent\textbf{Acknowledgements:} \\
The authors want to thank Andrea Dapor for collaboration in the early phases of the project and for many fruitful and helpful discussions on all aspects of this paper.\\
Further, we thank Thomas Thiemann, Benjamin Bahr and Laura Herold for several useful comments.\\
K.L. acknowledges support German National Merit Foundation as well as the German Research Foundation (DFG) under Germanys Excellence Strategy - EXC 2121 ``Quantum Universe'' - 390833306 and DFG-project BA 4966/1-2. Ł.R. would like to acknowledge
support by the Foundation for Polish Science
(IRAP project, ICTQT, Contract No. 2018/MAB/5, cofinanced
by the EU within the Smart Growth Operational
Programme).\\
Finally, K.L. thanks for many questions at the LOOPS '19 conference, where this project was first presented \cite{loops}.

\appendix

\section{General Relativity as SU(2) gauge theory and its kinematics on a graph}
The Hamiltonian or ADM formulation of general relativity \cite{ADM62} requires that one can split the four-dimensional manifold $\mathcal{M}$ in the following way: $\mathcal M \cong \re \times \sigma$ where $\sigma$ is a smooth 3-dimensional manifold admitting Riemannian metrics. Extending findings of Sen \cite{Sen82a}, it transpired that the ADM formulation can be casted into the form of a gauge theory with gauge group ${\rm SU}(2)$. The phase space is coordinatised by the Lie algebra valued connection $A_a(x)=A^I_a(x)\tau_I$ and the electric field $E^a(x)=E^a_I(x)\tau_I$ commonly called {\it Ashtekar-Barbero variables} \cite{Ash86,Bar94} with elementary Poisson brackets:
\begin{align}
\{E^a_I(x),E^b_J(y)\}=\{A^i_a(x),A^J_b(y)\}=0,\hspace{30pt}\{E^a_I(x),A^J_b(y)\}=\frac{\kappa\beta}{2}\delta^a_b\delta^J_I \delta^{(3)}(x,y),
\end{align}
where $\kappa=16\pi G$ is the gravitational coupling constant and $\beta\in \mathbb R -\{0\}$ is the Barbero-Immirzi parameter.\\
Moreover, the phase space is subject to the Gauss constraint 
\bal\label{GaussConst}
 G_J=\partial_a E^a_J+\epsilon_{JKL} A^K_a E^a_L =0,
\eal
in addition to the usual constraints of the ADM formulation, namely the {\it scalar constraint} $C$ and the {\it diffeomorphism constraint} $C^a$, which read in terms of the Ashtekar-Barbero variables: (assuming positive orientation of the triad from here on)
\bal
\label{diffeo_c}&D_a=\frac{2}{\kappa\beta}F^J_{ab}(A)E^b_J,\\
&\label{scalar_c}C=C_E-\frac{1+\beta^2}{\kappa}K_a^MK_b^N\epsilon_{MNJ}\epsilon^{JKL}\frac{E^a_KE^b_L}{\sqrt{\det{E}}},\\
&\label{scalar_ce}C_E:=\frac{1}{\kappa}F^J_{ab}(A)\epsilon_{JKL}\frac{E^a_KE^b_L}{\sqrt{\det(E)}}.
\eal
Here $K_a^I$ is the extrinsic curvature and $F_{ab}^I(A)=2\partial_{[a}A_{b]}^I+\epsilon_{IJK}A^J_aA^K_b$ is the curvature of the connection $A$.\\

Before proceeding to quantisation, we will have a look at discretisations of the spatial slice $\sigma$, in form of graphs $\gamma$ that allow an associated dual cell complex. Each graph is a collection of finitely many edges $e$, i.e. path in $\sigma$ and vertices $v$, i.e. ending points of multiple edges. For each $\gamma$ we will now introduce a phase space by considering the discretised phase space variables associated to $\gamma$ in the following sense: \\
Along the edges $e$ of the lattice, we will compute the holonomies $h(e)\in{\rm SU}(2)$ of the connection and along the associated faces $S_e$ of the dual cell complex (whose intersection with the edges we choose to be in the middle of each edge) we will compute the gauge covariant fluxes $P(S_e):=P^J(S_e)\tau_J$. For an edge $e$ these read: \cite{Thi00_gc}
\bal
h(e)&:=\mathcal{P}\exp\left(\int_0^1 dt\; A_a^J(e(t))\tau_J\dot{e}^a(t)\right)\label{holonomy},\\
P^J(e)&:=-2{\rm tr}(\tau_I h(e_{[0,1/2]})\int_{S_{e}} dx \; h(\rho_x) (\ast E)(x) h(\rho_x)^{\dagger} \; h(e_{[0,1/2]})^\dagger),\label{gaugecov_fluxes}
\eal
where in the path ordered exponential later vales are ordered to the right and $\rho_x$ is some, arbitrary path inside of $S_e$ such that $\rho_x(0)\in e_k$ and $\rho_x(1)=x$ and $e_{[0,1/2]}\subset e$ denotes the part of the path from $e[0]$ until $\rho_x(0)$.\\

Several comments are in order as the choice for the fluxes in (\ref{gaugecov_fluxes}) deviates from the standard one $E(e)=\int_{S_e} \ast E(x)$ which was used in \cite{DL17b}.\\
Firstly, upon taking a suitable continuum limit, in the sense of considering a family of graphs $\{\gamma_{\epsilon}\}_{\epsilon}$ that {\it fill out} $\sigma$ for $\epsilon\to 0$, then both $P(e):=P^J(e)\tau_J$ and $E(S_e)$ will reduce to $E^I_{a}(e[0])\dot{e}^a[0]$, i.e. to the electric field. Therefore, when considering the continuum limit, it does not make a difference which flux-regularisation to consider.\\
However, in presence of finite regularisations, i.e. for every finite graph $\gamma$, the situation is drastically different and favours the choice $P(e)$ in (\ref{gaugecov_fluxes}). For one, only this choice transforms covariantly under gauge transformations, i.e. under a gauge transformation $g(x)\in {\rm SU}(2)$ it is
\bal
P(e)\mapsto g(e[0])P(e)g(e[0])^{-1},
\eal
which allows to find regularisations of the constraints $D_a,C$ from (\ref{diffeo_c}) and (\ref{scalar_c}) such that they are ${\rm SU}(2)$ gauge invariant and similarly enables the construction of $SU(2)$ gauge invariant observables using only the basic building blocks (\ref{holonomy}) and (\ref{gaugecov_fluxes}) of the graph $\gamma$. If one wants to look at general relativity restricted to one single graph $\gamma$ it is therefore necessary to work with the gauge covariant fluxes in order to obtain physically meaningful results. And this will be the point of view taken in the remainder of this article.\\
Further, it is to be noted that the set of basic building blocks on a graph $\gamma$ form the phase space of a discretised theory only upon using fluxes, that do not have vanishing Poisson brackets, in order to avoid problems with the Jacobi identity, when (\ref{PB_HolFlu}) is assumed. As it is complicated to determine the Poisson brackets between the $E^I(e)$, the gauge covariant fluxes $P(e)$ are once again favoured as their Poisson bracket on $\gamma$ is uniquely found to be:
\bal
\label{PB_HolHol}\{h(e),h(e')\}_{\gamma}&=0,\\
\label{PB_HolFlu}\{P^I(e), h(e')\}_{\gamma}&=\frac{\kappa\beta}{2}\delta(e,e') \tau_I h(e),\\
\label{PB_FluFlu}\{P^I(e),P^J(e')\}_{\gamma}&=-\frac{\kappa\beta}{2}\delta(e,e') f^{IJ}_{\;K} P^K(e),
\eal
and obeys the Jacobi identity. Note, that $f^{IJ}_{\;K}$ are the structure functions of the chosen basis of $\mathfrak{su}(2)$ in the definition (\ref{gaugecov_fluxes}), i.e. $f^{IJ}_{\;K}=-[\tau_I]^{(1)}_{KJ}:=-{D'}^{(1)}_{KJ}(\tau_I)$ with $D'$ defined in \cite{DL17b}. For example, when using the basis $\uptau_I$ (on which we rely in section \ref{s3}) we have
\begin{align}\label{structure_function_tau}
[\uptau^{K}]_{ab}^{(j)}&
= i\sqrt{j(j+1)(2j+1)}(-1)^{j+b}\left(\begin{array}{ccc}
j & 1 & j\\
b & K & -a
\end{array}\right),
\end{align}
(note that the degenerate case $j=0$ is actually defined as $[\tau^K]^{(0)}\equiv 0$.)
and the $f^{IJ}_{\;K}$ become
\begin{align}
\left(\begin{array}{ccc}
f_{0+0} & f_{0++} & f_{0+-}\\
f_{0-0} & f_{0-+} & f_{0--}\\
f_{+-0} & f_{+-+} & f_{+--}
\end{array}
\right) = \left(
\begin{array}{ccc}
0 & -i & 0\\
0 & 0 & i\\
i & 0 & 0
\end{array}
\right)\; ,
\end{align}
(and antisymmetric under the first two indices), which by inspection of (\ref{structure_function_tau}) are closely related to
\begin{align}
\epsilon^{ijk}:=i\sqrt{6}\left(
\begin{array}{ccc}
1 & 1 & 1\\
i & j & k
\end{array}
\right)\;.
\end{align}

\end{document}